\documentstyle[12pt,aasms4]{article}
\def\refit{{\null}}
\def\parcsec{{\tt ''}\mskip -7.6mu.\,}   
\def\parcmin{{\tt '}\mskip -6.0mu.\,}    
\def\pdegree{^{\circ}\mskip -7.6mu.\,}   
\def\etal{{\it et~al.\ }}
\def\Halpha{{H$\alpha$\ }}
\def\simlt{\ {\raise-.5ex\hbox{$\buildrel<\over\sim$}}\ }
\def\kms{{km~s$^{-1}$}}
\def\ie{{\refit i.e.,}\ }
\def\eg{{\refit e.g.,\ }}
\def\etal{{\refit et al.\ }}
\def\apjlet{{\apj}}
\def\refit{{\null}}

\begin{document}
\title{The Planetary Nebula Luminosity Function of M87 and the 
Intracluster Stars of Virgo}

\author{Robin Ciardullo\altaffilmark{1,2}}
\affil{Department of Astronomy and Astrophysics, Penn State University,
525 Davey Lab, University Park, PA 16802}

\author{George H. Jacoby}
\affil{Kitt Peak National Observatory, National Optical Astronomy
Observatories, P.O. Box 26732, Tucson, AZ 85726}

\and

\author{John J. Feldmeier and Roger E. Bartlett}
\affil{Department of Astronomy and Astrophysics, Penn State University,
525 Davey Lab, University Park, PA 16802}

\altaffiltext{1} {Visiting Astronomer, Kitt Peak National Optical Astronomy
Observatories, which is operated by the Association of Universities for Research in Astronomy, Inc., under cooperative agreement with the National Science Foundation.} 

\altaffiltext{2}{NSF Young Investigator} 

\begin{abstract}
We present the results of a wide-field [O~III] $\lambda 5007$ survey for 
planetary nebulae (PN) in M87 and its surrounding halo.  In all, we identify 
338 PN candidates in a $16\arcmin \times 16\arcmin$ field around the galaxy;
187 of these objects are in a statistical sample which extends to 
$m_{5007} = 27.15$.  We show that the planetary nebula luminosity function 
(PNLF) of M87's halo is unlike any PNLF observed to date, with a shape
that differs from that of the empirical law at the 99.9\% confidence level.
In addition, we find that the PNLF of M87's outer halo differs from that of
the galaxy's inner regions at a high degree of certainty ($\sim 92\%$).  
We show that both these effects are most likely due to the existence of 
intracluster PN, many of which are foreground to M87.   These intracluster 
objects explain the ``overluminous'' [O~III] $\lambda 5007$ sources 
previously identified by Jacoby, Ciardullo, \& Ford (1990), and present us 
with a new tool with which to probe the morphological and dynamical 
properties of the cluster.

By modifying the maximum likelihood procedures of Ciardullo \etal (1989a)
to take into account the presence of ``field objects,'' and using an
assumed M31 distance of 770~Kpc (Freedman \& Madore 1990) with a
Burstein \& Heiles (1984) reddening law, we derive a distance
modulus to M87 of $30.79 \pm 0.16$ ($14.4 \pm 1.1$~Mpc).  This value is
in excellent agreement with the previous survey of Jacoby, Ciardullo, \& Ford 
(1990) and contradicts the assertion of Bottinelli \etal (1991) and Tammann 
(1992) that the PNLF distance to Virgo has been underestimated due to 
inadequate survey depth. 

\end{abstract}


\section{Introduction}
 
The planetary nebula luminosity function (PNLF) technique is one of the 
simplest methods for determining extragalactic distances out to
$\sim 20$~Mpc.  One takes a deep exposure of a galaxy through a filter
which passes light at [O~III] $\lambda 5007$, and compares the image
to a slightly deeper exposure off the emission line.  Those stellar objects 
that appear on the [O~III] $\lambda 5007$ image but not on the off-band frame
are either planetary nebulae (PN) or compact H~II regions.  If the target 
object is a normal elliptical or S0 galaxy with no star formation, then
the presence of H~II regions can generally be discounted, and one is left
with a list of PN, from which one can form a luminosity function.

The power of the PNLF technique comes from the shape of the luminosity
function.  At faint magnitudes, the PNLF has the power law form
predicted from models of uniformly expanding shells surrounding
slowly evolving central stars (Henize \& Westerlund 1963; Jacoby 1980).
However, observations have demonstrated that the bright end of the PN 
luminosity function dramatically breaks from this relation, and falls to
zero very quickly, within $\sim 0.7$~mag (cf.~Jacoby \etal 1992).  It is the 
constancy of this cutoff magnitude, $M^*$, and its high monochromatic
luminosity, that makes the PNLF such a useful standard candle. 

The shape and absolute magnitude of the PNLF cutoff has been successfully
reproduced theoretically by a number of authors, including Jacoby (1989), 
Dopita, Jacoby, \& Vassiliadis (1992), M\'endez \etal (1993), Han, 
Podsiadlowski, \& Eggleton (1994), and Richer, McCall, \& Arimoto (1997). 
Nevertheless, Bottinelli \etal (1991) and Tammann (1992) have argued that 
the bright-end of the PNLF is actually a power law, and thus observations
which do not reach the break in the luminosity function are not useful for 
distance measurements.  In support of this model, Bottinelli \etal and
Tammann point to the ``overluminous'' [O~III] $\lambda 5007$ sources found by 
Jacoby, Ciardullo, \& Ford  (1990) in Virgo elliptical galaxies, which can 
plausibly be argued to be part of a high-luminosity tail to the PNLF\null. 
By adopting the power-law model, and ignoring evidence for curvature in the 
observed PNLF of Virgo, Bottinelli \etal and Tammann have argued that the 
PNLF distance to this cluster is biased towards too low a value.  Although 
Kolmogorov-Smirnov and $\chi^2$ tests show this interpretation is highly 
unlikely, the most unambiguous way to test the hypothesis is to perform a 
deep, wide-field [O~III] $\lambda 5007$ imaging survey of the Virgo
ellipticals and better define the shape of the faint-end of the PNLF\null.

In this paper, we report the results of a $16\arcmin \times 16\arcmin$ 
[O~III] $\lambda 5007$ survey centered on the central elliptical of
Virgo, M87.  In \S 2, we give the details of the survey, present the 
positions and magnitudes of 338 planetary nebulae found in the galaxy's 
envelope and outer halo, and estimate the photometric accuracy of our 
measurements by comparing our derived PN magnitudes with those given in 
Jacoby, Ciardullo, \& Ford (1990).  In \S 3, we select two statistically 
complete subsets of these planetaries, and demonstrate that our planetary 
nebula luminosity function extends well onto the Henize \& Westerlund (1963) 
tail, making a distance determination to the galaxy possible.  In \S 4, we 
discuss the surprising result that the PNLF of M87's outer halo has a cutoff 
that is $\sim 0.2$~mag brighter than that for the inner part of the galaxy.  
We then show that, in retrospect, this behavior could have been predicted, 
since the intracluster stars of Virgo should produce a considerable number of 
planetary nebulae, and some of these objects will be foreground to M87{}.  
In \S 7, we include this ``field'' contribution in our maximum likelihood
analysis, and derive a distance to M87 of $14.4 \pm 1.1$~Mpc, in good
agreement with the previous PNLF distance determination to the galaxy.
This result vitiates the hypothesis of Bottinelli \etal (1991) and
Tammann (1992) that the bright-end of the PNLF is an unbounded power-law.   
We conclude by discussing the implications intracluster PN have for 
morphological and dynamical studies of nearby clusters.

\section {Observations and Reductions}
On 6 and 7 April 1995 we surveyed a $16\arcmin \times 16\arcmin$ region of
sky around M87 with the T2KB CCD on the Kitt Peak 4-m telescope, which
afforded a plate scale of $0\parcsec 47$ per pixel. Our on-band data 
consisted of seven exposures totaling 6.8~hours through a $\sim 30$~\AA\
full-width-half-maximum (FWHM) interference filter centered at 5030~\AA\ 
in the converging f/2.7 beam of the telescope.  (The transmission curve 
of this filter at the ambient temperature of $11^\circ$~C is displayed in
Figure~1.)  Our off-band data was composed of five 540~sec exposures through 
a 267~\AA\ wide filter centered at 5312~\AA.  In addition, an \Halpha image, 
consisting of nine 900~sec exposures through a 75~\AA\ FWHM interference 
filter centered at 6606~\AA, was obtained on 8 April 1995.  These latter 
data were used to discriminate PN from compact H~II regions, supernova 
remnants, and emission associated with M87's cooling flow.  The seeing for 
our $\lambda 5007$ on-band survey was $1\parcsec 2$; our \Halpha data was
taken in better than $1\parcsec 0$ seeing.

Planetary nebula candidates were identified and measured in a manner similar
to that described in detail by Jacoby \etal (1989), Ciardullo, Jacoby, \&
Ford (1989b), and Jacoby, Ciardullo, \& Ford (1990).  After spatially 
registering all the individual frames, we combined the on-band, off-band, 
and \Halpha frames of each field, using the {\tt imcombine} task in IRAF to 
reject radiation events.  We then  ``blinked'' the on-band [O~III] $\lambda 
5007$ sum against the off-band $\lambda 5312$ and \Halpha sum.  Objects 
clearly visible on the on-band image, but absent on the off-band and \Halpha 
frame were noted as possible planetaries.  We confirmed these identifications 
by examining the candidates on each individual on-band frame, and then 
looking closely at the appearance of each object on our [O~III] $\lambda
5007$ ``difference'' picture. 

Equatorial coordinates for the PN candidates were derived using 86 reference
stars from the USNO-A.1.0 Astrometric Catalog (Monet 1996) to define the 
CCD's coordinate system; the internal error in these coordinates is 
$\sim 0\parcsec 5$.  [O~III] $\lambda 5007$ photometry was accomplished 
relative to bright field stars with the DAOPHOT point-spread-function 
fitting routines (Stetson 1987) within IRAF\null.  These measurements were 
placed on the standard system by comparing large aperture measurements
of the field stars with similar measurements of the Stone (1977) and Oke 
(1974) spectrophotometric standards G191B2B, Feige~34, BD+25~3941, and 
BD+40~4032.  The dispersion in the photometric zero point computed from these 
stars was $0.03$~mag.  Finally, we computed the standard $\lambda 5007$ 
magnitudes for the PN by modeling the filter transmission curve (Jacoby \etal 
1989) and using the photometric procedures for emission-line objects
described by Jacoby, Quigley, \& Africano (1987).  For this computation, the
systemic velocity of M87 was taken from the Third Reference Catalog of Bright
Galaxies (de Vaucouleurs \etal 1991), and the galaxy's envelope velocity 
dispersion was estimated from Sargent \etal (1978).  Note that since the
systemic velocity of M87 is near the peak of the filter transmission curve, 
a $\sim 100$~\kms\ error in the latter quantity translates into a 
flux error of only $\sim 0.03$~mag.  

Table~1 lists the PN candidates identified in the field of M87, and follows
on from the numbering scheme of Jacoby, Ciardullo, \& Ford (1990).
Columns~2, 3, and 4 of the table list the objects' epoch 2000 coordinates and
$\lambda 5007$ magnitudes as defined by Ciardullo \etal (1989a),
\begin{equation}
m_{5007} = -2.5 {\rm \,log \,} F_{5007} - 13.74
\end{equation}
Column~5 gives the semi-major axis of the isophote upon which each PN is
superposed.  For $r_{\rm iso} < 5\parcmin 8$, these values were determined 
using the surface photometry of Cohen (1986); at larger distances, the 
isophotal radii were computed from an assumed axis ratio $b/a = 0.77$ and 
a galactic position angle of ${\rm p.a} = 158^\circ$.  Table~2 lists an 
additional 9 PN that are projected very near other galaxies in the field and 
are presumably bound to them.  These objects are included only for
completeness and are not used in any of our analyses.  

Table~3 gives the mean errors in our photometric measurements as reported
by the PSF-fitting algorithms of DAOPHOT\null.  However, because portions of 
M87 have been previously surveyed by Jacoby, Ciardullo, \& Ford (1990), 
it is possible to independently assess our errors by comparing the two
data sets.  Of the 55~PN identified by Jacoby {\refit et al.,} 45 were
recovered in this survey.  A comparison of the magnitudes of the four 
PN with the highest signal-to-noise ratio shows that there is no statistical
difference between the magnitude system of the two surveys: the zero point
of the new observations is $0.03 \pm 0.06$~mag brighter than that from 1990.
However, as Figure~2 demonstrates, a comparison of the entire dataset
indicates that there is an additional source of scatter $\sim 0.1$~mag above
that expected from the combined errors of the two measurements.  Part of the
scatter is probably due to differences in filter transmission curves, as
M87's internal velocity dispersion will shift the emission lines of some
objects on the filters' wings.  (We correct for this effect in the mean
using the techniques outlined in Jacoby \etal (1989) and Ciardullo,
Jacoby, \& Ford (1989b), but corrections for individual objects cannot
be made without velocity information.  Most of the additional error
probably comes from the 1990 data, which was compromised by variable seeing
and a high readout noise RCA CCD\null. Nevertheless, for the analysis
below, we have added an additional 0.07~mag uncertainty in quadrature to
the errors listed in Table~3.  In practice, the amplitude of the error
term makes very little difference to our final results.

Nine PN candidates from Jacoby, Ciardullo, \& Ford (1990) were not  
detected in this survey: PN \# 30, 35, 45, 49, 50, 51, 52, 53, and 54.
In addition, PN candidate \# 55 also was not recovered, but it fell at
the position of a CCD defect, and thus could not be checked.  Eight 
of these objects were at the limit of the previous survey and below
the stated completeness limit; the other two were near the limit of 
completeness.  All of the brighter PN from the previous survey were easily
recovered in this new, wide-field survey.

\section{The PNLF of M87's Halo}

Figure~3 displays the raw planetary nebula luminosity function for M87{}.
As is illustrated, our data extend well past the PNLF cutoff onto the
expected power-law tail.  Although these data are not statistically 
complete, and cannot be used for a precise distance estimate, it is
clear that the position of the PNLF cutoff is much brighter than the 
$m_{5007} \sim 27.0$ value needed if the Virgo core is at the distance
suggested by Sandage \& Tammann (1995, 1996).

To form a statistical sample of PN, we began by considering the 
detectability of planetary nebulae in our field.  For most of our
$16\arcmin \times 16\arcmin$ survey region, the background sky (which
is the dominant source of photometric noise) is constant, hence the
limiting magnitude for PN detections is constant.  However, near the
center of M87, the galaxy background dominates, and the detectability of 
faint PN decreases.  To address this problem, we used the results of 
Ciardullo \etal (1987) and Hui \etal (1993), who showed that PN detections 
are essentially 100\% complete when the DAOPHOT signal-to-noise ratio (SNR)
is greater than $\sim 9$, but no PN are detected when the SNR is less than
$\sim 4$.  Thus, for our statistical sample, we included only those PN with 
a DAOPHOT measurement error of $\sigma_{\rm err} < 1.086 / {\rm SNR} = 
0.12$~mag.  At isophotal galactic radii greater than $2\arcmin$, this 
completeness criteria included essentially all PN brighter than $m_{5007} 
= 27.2$.  At galactic radii smaller than this, however, the limiting
magnitude decreased quickly, so that by $r_{\rm iso} < 0\parcmin 5$, no PN
were detectable.  We therefore defined our ``complete sample'' to be those
PN with $m_{5007} < 27.15$ and $r_{\rm iso} > 2\arcmin$. 

\section{The Planetary Nebula Luminosity Function of M87}
Figure~4 plots the planetary nebula luminosity function for the statistical
sample of PN{}.  From these data, the PNLF distance to the galaxy can
normally be derived by convolving the empirical model for the PNLF given
by Ciardullo \etal (1989a)
\begin{equation}
N(M) \propto e^{0.307M} \, [1 - e^{3(M^{*}-M)}]
\end{equation}
with the photometric error function (Table~3) and fitting the data to the 
resultant curve via the method of maximum likelihood.  However, in the
case of M87, the most likely empirical curve is a poor fit to the luminosity
function (cf.~Figure~4).  In fact, Kolmogorov-Smirnov and $\chi^2$ tests both 
exclude the Ciardullo \etal law at the 99.9\% confidence interval.  This is 
a remarkable result: none of the PNLFs from any of the $\sim 30$ previously 
studied galaxies differs significantly from the empirical law.  Moreover, 
the large number of PN detected in this survey cannot be invoked to explain 
the discrepancy.  The luminosity functions of M31 (Ciardullo \etal 1989a), 
M81 (Jacoby \etal 1989), NGC~5128 (Hui \etal 1993), and NGC~4494 (Jacoby,
Ciardullo, \& Harris 1996) all contain similar numbers of objects.
The planetaries surrounding M87 are therefore unique in some way.

An even more surprising result comes if we divide our PN sample 
in two, and compare the PNLFs of M87's inner and outer halo.  For the
inner sample (sample ``A''), we take all the PN in our statistical sample 
with isophotal radii between $2\arcmin$ and $4\arcmin$; for the outer
sample, (sample ``B'') we take those PN with $r_{\rm iso} > 4\arcmin$.  
Both samples are plotted in Figure~5.  As is illustrated, sample ``A''
contains PN \#1, the extremely overluminous object first identified by 
Jacoby, Ciardullo, \& Ford (1990).  However, of the 20 brightest PN in the
samples, 18 belong to sample ``B''.  More importantly, the shapes of the 
two PNLFs appear different: a Kolmogorov-Smirnov test reveals that the two 
samples are different at the 92\% confidence level.  Again, this result is 
unique.  Explicit tests for changes in the PNLF cutoff with galactocentric 
radius have been performed with the large samples of PN available in NGC~5128 
(224 objects; Hui \etal 1993) and NGC~4494 (101 objects; Jacoby, Ciardullo, 
\& Harris 1996).  In neither case was a gradient observed.  

\section{Explaining the Luminosity Function}
Internal and external tests on the $\sim 30$ early and late-type galaxies 
surveyed to date have shown that the PNLF cutoff is remarkably insensitive to
changes in stellar population (cf.~Jacoby 1996).  However, a number
of mechanisms do exist which can, at least in theory, cause the PNLF
technique to fail and produce a change in the observed value of $m^*$.  
The first, and simplest, is to hypothesize that some instrumental effect 
exists, such as a radial gradient in the flatfield or the transmission curve 
of the filter.  We have examined the former possibility by comparing the
flatfields taken through our [O~III] $\lambda 5007$ Virgo filter with similar
flats taken through a different (lower redshift) [O~III] $\lambda 5007$
filter that same night (cf.~Feldmeier, Ciardullo, \& Jacoby 1997).  No
large-scale difference is apparent.  Similarly, we tested for problems with
the filter bandpass by tracing the transmission curve of our filter at four
different positions on the glass: to within $\sim 3$~\AA, the central
wavelength of our filter is the same at all points.  Since at the velocity
of M87, a $\sim 3$~\AA\ shift in wavelength corresponds to, at most, a
$\sim 3\%$ change in the filter transmission (cf.~Figure~1), the effect is
much too small to explain our result.

A second method for explaining the variation in $m^*$ is to invoke 
non-uniform extinction in the galaxy.  Dust has been detected in the central
regions of many elliptical galaxies (\eg Sadler \& Gerhard 1985;
Ebneter, Djorgovski, \& Davis 1988; Goudfrooij \etal 1994), and several
authors have suggested that extinction is responsible for the presence of 
radial color gradients in these systems (Witt, Thronson, \& Capuano 1992;
Goudfrooij \& De Jong 1995; Wise \& Silva 1996).  However, these
studies have concentrated on the central regions of elliptical galaxies,
while our survey deals exclusively with PN that are more than 1.5 effective 
radii ($r_e$) from the galactic nucleus.  In fact, there is little reason to 
believe that the extinction in M87 changes by $E(B$$-$$V) \sim 0.05$ between 
2 and $4 \, r_e$.  Moreover, even if there is a strong gradient in the 
dust distribution, this still may not translate into an observed gradient in
$m^*$.  As Feldmeier, Ciardullo, \& Jacoby (1997) have shown, the location
of a galaxy's PNLF cutoff is relatively insensitive to the presence of
dust, as long as the scale length for the obscuration is smaller than that
of the stars.  (This is because the bright end of the PNLF (\ie the PNLF
cutoff) is always dominated by the bright, unextincted PN{}.  Unless the
dust extends far enough to cover virtually all the PN, the bright-end
cutoff will always consist of unreddened objects.)  This makes it very
unlikely that dust is responsible for the change in $m^*$.
  
Although the absolute magnitude of the PNLF cutoff is extremely 
insensitive to the details of the underlying stellar population, a dramatic
change in the metallicity or age of M87's halo stars could, in principle,
produce a change in $m^*$ similar to that observed.  Both observations 
(Ciardullo \& Jacoby 1992; Richer 1994) and theory (Dopita, Jacoby \& 
Vassiliadis 1992) suggest that galaxies with [O/H] $\simlt -0.5$ can have 
a PNLF cutoff that is different from that of metal-rich populations by 
$\sim 0.1$~mag.  Unfortunately, this effect acts in the wrong direction:
it is the metal-poor systems that have fainter values of $M^*$.  In 
order to explain the observed gradient, the center of M87 would have to
be metal-poor, and the halo would need to be metal-rich.  Observations
show that this is extremely unlikely (Kormendy \& Djorgovski 1989).

Similarly, it is difficult to use population age to explain PNLF variations.
According to the models of Dopita, Jacoby, \& Vassiliadis (1992) and M\'endez 
\etal (1993), the location of the PNLF cutoff is nearly independent of age
for populations between 3 and 12~Gyr.  If these models are correct, then in
order to enhance the luminosity of the PNLF cutoff in M87's outer halo, one
must hypothesize an unrealistically young age for the stars, $\sim 0.5$~Gyr.  
Moreover, even this may not be sufficient, as there is excellent agreement 
between the PNLF and Cepheid distances to the Large Magellanic Cloud and M101
(Jacoby, Walker, \& Ciardullo 1990; Feldmeier, Ciardullo, \& Jacoby 1997). 
These observations strongly suggest that the location of the PNLF cutoff 
does not change by much, even in systems with active star formation.  It is
therefore difficult to attribute changes in the PNLF cutoff to variations in
stellar population.  

\section{Planetary Nebulae and the Intracluster Stars of Virgo}
The best hope for explaining the observed changes in M87's halo PNLF lies
in the Virgo Cluster itself.  All of the above explanations implicitly
assume that the PN projected onto M87's outer halo are at the same
distance as those PN which are members of the inner sample.  However, if the 
Virgo Cluster has a substantial population of intracluster stars, this will 
not be the case, as some objects will be superposed in the foreground, and
others will be in the background.  For example, if the Virgo Cluster is at a
distance of $\sim 15$~Mpc, then the central $6^\circ$ core of the cluster
(de Vaucouleurs 1961), has a linear extent of $\sim 1.5$~Mpc.  If the core
is spherically symmetric and filled with stars, then we might expect some 
intracluster objects to be up to $\sim 0.25$~mag brighter than the value of 
$m^*$ derived from galaxies at the center of the cluster.  This is roughly
what is observed in Figure~5.

Further evidence that the anomalous PNLF of M87 is due to foreground 
contamination comes from the fact that it is the outer sample of objects
that has most of the bright PN{}.  The number of foreground PN detected 
in any region of our CCD field should be roughly proportional to the area 
of the field; since the outer region samples $\sim 8$ times more area than 
the inner field, those data should contain $\sim 8$ times more intracluster
objects.  In addition, M87's sharply peaked surface brightness profile
guarantees that the ratio of galaxy light to intracluster light in the
inner field will be much larger than that in the outer field.  Consequently,
the contribution of intracluster objects to sample ``A'' will be small,
while that for sample ``B'' will be relatively large.  Again, this is
roughly what is displayed in Figure~5.

The intracluster PN hypothesis also explains the existence of
``overluminous'' planetary nebulae.  In their survey of Virgo, Jacoby,
Ciardullo, \& Ford (1990) identified a small number of [O~III] $\lambda 5007$ 
sources that were significantly brighter than predicted by the empirical
planetary nebula luminosity function.  As discussed by Jacoby, Ciardullo,
\& Harris (1996), there are a number of possible origins for these objects,
including compact H~II regions, supernova remnants, Wolf-Rayet nebulae,
supersoft x-ray nebulae, coalesced binary-star nebulae, chance superpositions
of multiple objects, and even background quasars at $z = 3.1$.   However,
most of these explanations are not satisfactory from a stellar population
standpoint, and the remaining ideas have little observational support.  In
fact, to investigate this question, we imaged the overluminous object
NGC~4406 PN \#1 (Jacoby, Ciardullo, \& Ford 1990) with the Planetary Camera
of the {\sl Hubble Space Telescope\/}.  Even at WFPC~II resolution, the
object is unresolved.  Since $\sim 0\parcsec 05$ at Virgo corresponds to a
linear size of $\sim 4$~pc, this observation effectively excludes H~II
regions, supernova remnants, and PN superpositions from the list of
possibilities.

The existence of intracluster PN provides a natural explanation for
the overluminous planetaries.  It also explains why similarly bright
[O~III] $\lambda 5007$ sources have not been detected in the Leo~I,
Triangulum, Coma~I, or Fornax Clusters, nor in any field galaxy.  If 
the overluminous PN are indeed intracluster in origin, then isolated field
galaxies will be clean systems without foreground contaminants.  Furthermore, 
of the clusters observed to date, only Fornax is rich enough to have a 
substantial intracluster environment, but it has very little front-to-back 
depth (Tonry 1991; McMillan, Ciardullo, \& Jacoby 1993).  Virgo is the only 
cluster surveyed for PN where depth of field is important, and it is the
only cluster in which overluminous planetaries are found.

Although the idea of intracluster planetary nebulae seems speculative, there 
is, in fact, conclusive evidence that such objects do exist.   In their
radial velocity survey of 19~PN in the halo of the Virgo Cluster elliptical
M86 ($v = -220$~\kms), Arnaboldi \etal (1996) found three objects with
$v > 1300$~\kms.  These planetaries are undoubtably intracluster in origin.  
Significantly, the PN observed by Arnaboldi \etal were originally identified 
with a 30~\AA\ filter centered at 5007~\AA\ (Jacoby, Ciardullo, \& Ford
1990); intracluster objects with $v > 1000$~\kms\ should have been strongly
excluded.  The only reason three were detected was that at $v \sim 1500$~\kms,
[O~III] $\lambda 4959$ is shifted into the bandpass of the [O~III] $\lambda
5007$ filter!  Since $I(\lambda 5007)/I(\lambda 4959) = 3$, only the very
brightest PN could have been detected in this way.  The existence of three
intracluster objects in the Arnaboldi \etal sample therefore implies the
presence of many more.

Similar evidence for a population of intragalactic stars in Virgo comes from 
SN~1980I.  This event occurred in intergalactic space, midway between 
the Virgo core ellipticals NGC~4374 and NGC~4406.  Considering that only 
$\sim 12$ probable SN~Ia supernovae have occurred in the cluster core 
this century (Barbon, Cappallaro, \& Turatto 1989), the existence of one 
intergalactic object suggests that intracluster stars must be fairly 
common.

Intracluster PN have also been proven to exist in the Fornax Cluster.
An imaging survey of three intergalactic fields by Theuns \& Warren (1997) 
turned up ten planetary nebula candidates in 104 sq.~arcmin of sky.
When this detection rate is extrapolated to the entire cluster, 
the result is that a substantial fraction of the Fornax Cluster's stars,
perhaps as much as $\sim 40$\%, must be in intergalactic space.  If the same 
is true for Virgo, then the effect of intracluster objects on the galactic
PNLFs cannot be ignored.

In fact, a quantitative estimate for the stellar mass of Virgo's intracluster 
environment comes from observations of faint stars by the {\sl Hubble Space
Telescope.}  By comparing the number counts of objects in a Virgo field
with that in the Hubble Deep Field, Ferguson, Tanvir, \& Von Hippel (1997)
found an excess of $\sim 600$ stars in a $4.6$ square arcmin region.  If
these are red giant and asymptotic giant branch stars, and if their mean
lifetime at this luminosity is $\sim 10^6$~years, then a simple scaling of 
areas and lifetimes implies the existence of $\sim 700$~PN in our survey
field.  If, as implied by equation (2), $\sim 2\%$ of these PN are bright
enough to be observable, then our PNLF of M87 should be contaminated by
$\sim 14$ intracluster objects.  Given the uncertainties involved, this is
in agreement with our data.

\section{The Distance to M87}
The presence of intragalactic planetary nebulae in the Virgo Cluster 
impedes our ability to measure M87's distance via the planetary nebula
luminosity function technique.  Of the two samples of PN plotted in Figure~5, 
clearly sample ``A'' is the better one to use for this purpose, but
even it has some contamination from intracluster objects.  This ``field''
contribution must be removed before any distance measurement to the galaxy
can be attempted.

To do this, we first made the assumption that all PN with isophotal radii 
$r > 4 \parcmin 8$ ($r > 3 r_e$) are intracluster in origin.  (This is 
probably not strictly true; bound PN have been found $\sim 3.7~r_e$ from 
the center of NGC~3379 (Ciardullo, Jacoby, \& Dejonghe 1993) and more than 
$\sim 4 r_e$ away from the nucleus in NGC~5128 (Hui \etal 1995).  However,
this radius represents the best compromise between the need to exclude
galactic objects and the desire for good counting statistics.)  We then 
binned our sample of intracluster PN (sample ``C'') into 0.2~mag intervals, 
and fit a smooth curve through the points.  This curve, displayed in
Figure~6, represents our estimate of the intracluster PN luminosity function.

With the field luminosity function in hand, we proceeded to derive an 
expression for the most likely distance of M87{}.  If we treat both the
empirical PNLF of equation (2) and the intracluster PNLF as probability
distributions, then in any magnitude range, $dm$, the expected number of
observable PN is 
\begin{equation}
\lambda(m) dm = \left[ T \, \epsilon \circ \phi(m - \mu) + \xi \beta(m) 
\right] dm 
\end{equation}
where $\epsilon \circ \phi(M)$ is the convolution of the photometric
error function (Table~3) with the (normalized) empirical luminosity function, 
$\beta(m)$ is the (normalized) intracluster luminosity function, $T$ is the
total population of M87 planetaries in our field, and $\xi$ is the total
number of intracluster PN in our field.  Following Hanes \& Whittaker (1987) 
and Ciardullo \etal (1989a), the relative probability of observing a given 
set of $N$ planetaries down to limiting magnitude $m_l$, as a function of 
distance modulus $\mu$, $T$, and $\xi$, is
\begin{equation}
\ln P(T,\mu,\xi) = - \int_{-\infty}^{m_l} T \phi(m - \mu) dm - 
\int_{-\infty}^{m_l} \xi \beta(m) dm + \sum_i^N \ln 
\left\{ T \phi(m - \mu) + \xi \beta(m) \right\}
\end{equation}
Now note that, although $\xi$ is an independent variable, it is constrained
by the observations.  If $N_f$ is the number of intracluster PN in sample
``C'', and $A$ is the ratio of the area of sample ''A'' to the area of sample
''C'', then $\xi$ should be distributed according to the Poisson distribution
function, $P_P$, with a mean of $N_f \, A = 11.1$ and a standard  deviation
of $\sqrt{N_f} \, A = 1.25$.  Thus, when integrated over all $T$ and $\mu$, 
equation (4) should give
\begin{equation}
\int \int P(T, \mu, \xi) \, dT \, d\mu = P_P(\xi)
\end{equation}
This renormalization of (4) enabled us to derive a distance to M87
despite the fact that the functions $\phi(M)$ and $\beta(m)$ are somewhat
similar in shape.  The formulation also maintains the advantage of the
original PNLF maximum-likelihood technique, in that it avoids all 
problems associated with the binning of small amounts of data.

Figure~7 displays the likelihood of each solution as a function of M87's
distance modulus.  For the plot, we have assumed a foreground Galactic
extinction of $E(B$$-$$V) = 0.023$ (Burstein \& Heiles (1984), a Seaton
(1979) reddening law, and a value of $M^* = -4.54$ based on an M31 distance
of 770~kpc (Freedman \& Madore 1990) and an M31 extinction of
$E(B$$-$$V)_{\rm M31} = 0.08$ (Burstein \& Heiles 1984).  The most likely
distance modulus is $(m-M)_0 = 30.79$ (14.4~Mpc); the formal uncertainty of
the result is $+0.07$~mag, $-0.06$~mag. This distance would decrease by
$\sim 3\%$ if the Ciardullo \etal (1989a) value of $M^*$ ($-4.48$) were used. 

The present result is in excellent agreement with the previous PNLF 
distance determination.  When scaled to the same set of assumptions regarding
the PNLF zero point and Galactic extinction, the Jacoby, Ciardullo, \& Ford 
(1990) value for M87 is $(m-M)_0 = 30.85 \pm 0.09$ ($14.8 \pm 0.6$~Mpc).
Note that the previous survey was limited by a small sample size (only
36~objects in the statistical sample) and a noisy CCD chip.  It did, however,
extend to smaller galactocentric radii than the present survey, and thus
was not as badly contaminated by intracluster objects.  Consequently,
although the present survey contains many more PN and has a greater
photometric accuracy, the distance derivation is not significantly better
than the older work.

An interesting feature of our new PNLF distance measurement is the 
robustness of the result.  Because the empirical PNLF of equation (2)
goes to zero at $m < m^*$, the earlier PNLF measurements in Virgo
were sensitive to the luminosity of the brightest one or two planetaries.
In fact, as Jacoby, Ciardullo, \& Ford (1990) pointed out, the inclusion 
(or exclusion) of overluminous objects from their Virgo samples could
change the derived distances to galaxies by more than $\sim 20\%$.  (Jacoby
\etal handled this problem by noting that overluminous PN could be excluded
based on the overall quality of the fits --- solutions that included
overluminous objects were significantly poorer than fits without them.)
The present analysis, however, is much less sensitive to the existence of
these objects.  For example, if PN \#1 is arbitrarily excluded from our 
sample, the derived distance to M87 changes by less than 0.02~mag. Figure~6 
plots the best fit luminosity function, $T \phi(m - \mu) + \xi \beta(m)$.
As is illustrated, the fit is excellent.

The agreement between the two PNLF distance estimates strongly refutes
the contention of Bottinelli \etal (1991) and Tammann (1992) that sample
size affects PNLF distance measurements.  The new measurement uses over 
a factor of two more planetaries than the older determination, yet the 
distance to the galaxy remains substantially the same.  Moreover, the 
presence of intracluster stars in Virgo negates one of the core tenets
of Tammann (1992), that overluminous PN are a natural extension of the 
empirical PNLF{}.  This interpretation is no longer needed to explain the 
data.

The distance error quoted above is the formal uncertainty derived from
our maximum likelihood procedure, not the total error in the computed
distance.  To obtain the latter quantity, the uncertainty in the solution
must first be combined with usual random uncertainties associated with the
photometric zero point (0.03~mag), the correction associated with going from
the DAOPHOT magnitudes to the large aperture magnitudes (0.02~mag), the
filter calibration (0.03~mag), and the Galactic extinction (0.05~mag).  In
addition, the present analysis has an additional error, that associated with
the uncertain {\it shape\/} of the field PN luminosity function.  Simulations
suggest that this error is small, probably less than $\sim 0.05$~mag.  
Thus, the total random uncertainty in our distance determination is $\sim 
0.11$~mag.

Finally, two sources of error that affect all PNLF measurements is the
uncertainty in the distance to M31, which provides the zero point
(0.10~mag), and the definition of the empirical PNLF itself (0.05~mag).  
The addition of these two quantities in quadrature yields a total
uncertainty of 0.16~mag.

Our new distance modulus to M87, $(m-M)_0 = 30.79 \pm 0.16$ ($14.4 \pm 
1.1$~Mpc) is consistent with distances to the galaxy derived from the
surface brightness fluctuation method ($30.92 \pm 0.12$; Ciardullo, Jacoby, 
\& Tonry 1993, Tonry 1997), and the globular cluster luminosity function
($31.12 \pm 0.26$; Whitmore \etal 1995).  It is, however, marginally smaller
than the median distance of $16.1 \pm 1.0$~Mpc obtained from Cepheid 
measurements in the Virgo Cluster (Pierce \etal 1994; Ferrarese \etal 1996; 
Sandage \etal 1996; Saha \etal 1996a,b).  This difference may not be
significant; our results point out the problem with using Cepheids to 
estimate the distance to the Virgo Cluster core.  The brightest PN in 
sample  ``C'' is $\sim 0.35$~mag brighter than the value of $m^*$ in M87.
This implies that the Virgo Cluster {\it core\/} extends at least 2.1~Mpc
in depth.  Furthermore, of the five spirals with Cepheid measurements, only  
three are actually projected within the $6^\circ\,$ core, and only NGC~4571
(which has a ground-based Cepheid distance of $14.9 \pm 1.2$~Mpc) shows any
evidence of being physically {\it in\/} the core (cf.~van der Hulst \etal
1987).  Thus, the issue of cluster depth cannot be neglected.

Nevertheless, our PNLF measurements should give a reliable estimate of the
distance to the Virgo core.  Although M87 is offset by about $1^\circ$ from
the center of the cluster's isopleths, and has a radial velocity that is
$\sim 200$~\kms\ larger than the cluster mean (Binggeli, Tammann, \& Sandage
1987), the galaxy is almost certainly at the center of the cluster's
potential.  Velocity and positional offsets, such as those observed for M87,
are common in cD galaxies within dynamically young clusters (Bird 1994), and
there is good evidence to suggest that Virgo is such a cluster.  Based on
the radial velocities of dwarf galaxies inside the Virgo Cluster core,
Binggeli, Popescu, \& Tammann (1993) have concluded that there is a separate
group of galaxies associated with M86 (NGC~4406) that is falling into Virgo
from the far side of the cluster.  Independent confirmation of this
hypothesis comes from Jacoby, Ciardullo, \& Ford (1990), whose PNLF distances
place M86 (radial velocity $-220$~\kms) over 1~Mpc behind M87.  If this
interpretation is correct, then the masses of M86 and M87, as derived
from their x-ray halos, imply that the peculiar position and velocity of M87
is a direct consequence of the disturbance caused by the infall of M86 
and its group (B\"ohringer \etal 1994).

In fact, the best evidence of M87's position within the Virgo Cluster
comes from the x-ray data obtained by the {\sl ROSAT\/} satellite.  The 
distribution of Virgo's intracluster x-ray gas demonstrates quite 
conclusively that a large portion of the cluster's mass is centered at the 
position M87 (B\"ohringer \etal 1994).  This fact, along with the presence
of a diffuse cD halo around the galaxy (cf.~Weil, Bland-Hawthorn, \& Malin
1997), strongly indicates that M87 is, indeed, at the center of the cluster.
Our distance measurement to M87 should therefore be representative of that
of the group.

\section{Planetary Nebulae as Probes of the Virgo Cluster}
The existence of intracluster planetary nebulae provides us with a new and 
unique tool for probing the structure of Virgo.  For example, Huchra 
(1985, 1988) has pointed out that, from isopleths and velocity measurements 
alone, it is difficult to tell whether or not the central $6^\circ$ core of 
Virgo is virialized.  Evidence for virialization includes the nearly Gaussian 
distribution of velocities for the early-type galaxies, and the excellent 
agreement between the apparent morphology of the core and distribution of
galaxies expected from an isothermal King (1962) model; evidence against 
virialization comes from the positional and velocity offset between the
x-ray halo of M87 and the cluster center.  Thus, while Huchra (1988) states 
that the Virgo ellipticals form an apparently virialized core, Binggeli, 
Tammann, \&  Sandage (1987), and Binggeli, Popescu, \& Tammann (1993) 
argue strongly that the cluster contains significant substructure, even near 
its center. 

The observed planetary nebula luminosity function can help answer this
question.  Old stellar populations produce planetary nebulae in
proportion to the population's total luminosity (Renzini \& Buzzoni 1986; 
Ciardullo 1995).  Hence the number of intergalactic PN detected in our survey 
should reflect the total amount of intracluster luminosity present in our 
field.  More importantly, since the observed PNLF is formed from a
superposition of PNLFs at different distances along the line-of-sight, the 
precise shape of the observed luminosity function contains information on the
cluster's front-to-back morphology.  It is therefore possible to test models
of the Virgo Cluster by comparing the field PNLF of Figure~6 with the
luminosity distributions expected from different cluster profiles.

Figure~8 demonstrates this possibility.   In the figure, we compare the field
PNLF of M87 (sample ``C'') with the PNLF expected from an isothermal cluster
with a core radius of $r_c = 0.5$~Mpc and a center $52\arcmin$ from the
position of our survey field (Binggeli, Tammann, \& Sandage 1987).  It
is obvious that the model fails to fit the bright end of the PNLF by many
orders of magnitude.  In fact, this result is a general property of all
dynamically relaxed cluster profiles: once virialization occurs, the cluster
becomes much too condensed to explain the large number of (presumably)
foreground PN present in the observed PNLF\null.  To come close to producing
the requisite number of bright objects, the radial profile of the cluster
must be shallower.  However, as Figure~8 shows, even a uniform density law
fails to fit the overall shape of observed PNLF\null.  Consequently, it is
likely that the distribution of intergalactic stars in Virgo is clumpy, and
that the cluster is not virialized.   

Since planetary nebulae sample the light, it is possible, at least in
theory, to use our observations to place a constraint on the fraction of
the Virgo Cluster's dark matter that is in intergalactic stars.  In
practice, however, three difficulties prevent us from using our data in
this manner.  First is the limitation imposed by our narrow-band filter.
The interference filter used for this project was designed to detect the
PN associated with M87, and its 30~\AA\  FWHM bandpass is ideal for this
project (cf.~Figure~1).  However, the Virgo Cluster core has an observed
velocity dispersion of $\sigma_v \approx 800$~\kms\ (Binggeli, Tammann, \& 
Sandage 1987), or $\sim 26$~\AA\ at the wavelength of the [O~III] 
$\lambda 5007$ emission line.  As a result, our survey is probably missing
a significant number of intracluster planetaries, perhaps as many as 30\%.

A second difficulty with the interpretation of our data is that our CCD
field is not at a ``typical'' location in the Virgo Cluster, but is instead
centered on M87.  Consequently, the density of intracluster objects in our
field is likely to be higher than average.   Moreover, we cannot exclude
the possibility that some of the PN at large galactocentric radii are
actually bound to the M87.  While it may be possible to remove this
contribution by extrapolating the luminosity profile of M87 out to
large radii, this profile is not very well known (cf.~Weil, Bland-Hawthorn,
\& Malin 1997; Graham \etal 1996).  Furthermore, any extrapolation would
require the extra assumption that M87's luminosity-specific PN number density
does not change with radius.  Unfortunately, this quantity has been observed 
to increase with galactocentric radius in NGC~5128 (Hui \etal 1993) and in a 
composite galaxy formed from eight elliptical galaxies (Ciardullo, Jacoby, 
\& Feldmeier 1995).  Thus, the validity of this assumption is unknown.

By far the most serious obstacle encountered when trying to use our PN 
detections to estimate the density of Virgo's intracluster stars comes from 
the finite depth of the cluster and the steeply rising bright-end of the 
PNLF\null.  Our [O~III] $\lambda 5007$ survey reached a limiting magnitude of 
$m_{5007} = 27.15$; thus, if the distance to the Virgo Cluster core is 
$\sim 15$~Mpc and its size is $\sim 1.5$~Mpc, then our measurements sample
the top $\sim 1$~mag of the PNLF on the front side of the cluster.  On the
back-side of Virgo, however, our [O~III] $\lambda 5007$ measurements only go 
$\sim 0.5$~mag down the luminosity function.  In terms of equation (2), 
this means that the intracluster luminosity in the foreground of Virgo 
receives $\sim 2.6$ times more weight than luminosity on the backside.  If
the radial profile of the Virgo Cluster were smooth, this effect, as well
as the others presented above, could be modeled.  However, since it is
likely that sub-structure is present in the core of Virgo, any extrapolation
of the results from our one CCD field to the entire cluster is premature.

\section{Conclusion}
We have detected 338 planetary nebulae in our new, wide-field survey
of M87 and its surrounding halo.  The analysis of the luminosity function
of these PN demonstrates that M87 is at a distance of $14.4 \pm 1.1$~Mpc;
this number is in excellent agreement with the earlier PNLF measurement,
as well as recent distance determinations from the surface brightness
fluctuation method and the globular cluster luminosity function.  The
result is, however, in sharp disagreement with the hypothesis of
Bottinelli \etal (1991) and Tammann (1992) that PNLF measurements in
Virgo are biased due to a limited sample size.

Our observations of M87 also present strong evidence for the presence
of a substantial population of intergalactic stars, which extends over
$\sim 2$~Mpc in front of M87.  The intergalactic stars are, in all
likelihood, the explanation for the ``overluminous'' planetaries 
encountered by Jacoby, Ciardullo, \& Ford (1990) in Virgo, but not seen 
any where else (Jacoby, Ciardullo \& Harris 1996).  The analysis of the
intracluster PN luminosity function suggests that the Virgo Cluster core
is not virialized, but is instead dynamically young.  However, the mismatch
between the width of our [O~III] $\lambda 5007$ filter and the Virgo Cluster
velocity dispersion precludes a definitive statement or a more detailed
analysis at the present time.

Planetary nebula observations offer a new opportunity for morphological
and dynamical studies of nearby clusters.   To date, the only way to 
study intergalactic light in clusters has been through deep surface
photometry, and as a result, only a few, very rich, Abell Clusters have been
measured (cf.~Uson, Boughn, \& Kuhn 1991; V\'ilchez-G\'omez, Pell\'o, \&
Sanahuja 1994).  Planetary nebula observations offer an alternative to these
studies, and provide information on both the two and three dimensional
structure of the cluster.  Moreover, a PN's [O~III] $\lambda 5007$ emission
line is an excellent target for a radial velocity measurement.  Since there
are never enough sufficiently bright galaxies in a cluster to fully define
the cluster's velocity field, planetary nebulae can provide invaluable data
for cluster dynamics.  Future [O~III] $\lambda 5007$ surveys may therefore
allow us to probe the effects of galactic mergers, cluster accretion, and 
tidal-stripping within several nearby clusters, and enable new investigations
into the distribution of dark matter in clusters and of the initial 
conditions of cluster formation.

We would like to thank Mike Pierce for first suggesting the possible
intracluster origin of some of M87's halo PN\null.  This work was supported
in part by NASA grant NAGW-3159 and NSF grant AST-9529270.  Support for this
work was also partially provided by NASA through grant number GO-0612.01-94A
from the Space Telescope Science Institute, which is operated by the
Association of Universities for Research in Astronomy, Inc., under NASA
contract NAS5-26555.

\clearpage

\figcaption[] {The transmission curve of our [O~III] $\lambda 5007$ filter
when placed in the f/2.7 beam of the Kitt Peak 4-m telescope at the outside  
temperature of $11^\circ$~C.  Also shown is the $\pm 2 \, \sigma$ velocity
dispersion of M87's envelope (from Sargent \etal 1978) and the $\pm 2 \, 
\sigma$ velocity dispersion of the Virgo Cluster (from Binggeli, Tammann, \&
Sandage 1987).}

\figcaption[] {A comparison of the PN [O~III] $\lambda 5007$ magnitudes
of this survey with those obtain by Jacoby, Ciardullo, \& Ford (1990).  The
dotted lines define the internal $1 \, \sigma$ photometric error derived
in DAOPHOT{}.  Although the data suggest that DAOPHOT underestimates the
true photometric error by as much as $\sim 0.1$, some of the additional 
scatter can be attributed to differences in the transmission curves.
Most of the outliers lie on areas are projected on areas of high galaxy
background, where the uncertainty in determining the sky level dominates.}

\figcaption[]  {The luminosity function for our entire set of planetary 
nebulae around M87 binned into $0.1$~mag intervals.  Although the data do not 
represent a statistical sample, it is obvious that the data reach past the
cutoff to the power-law tail of the faint-end of the luminosity function.  
It is also clear that for $M^* = -4.5$, the distance modulus of $(m - M) 
\sim 31.5$ advocated by Sandage \& Tammann (1995; 1996) is incompatible
with the data.}

\figcaption[]  {The planetary nebula luminosity function for a sample of
planetaries with isophotal radii greater than $2\arcmin$ binned into 
$0.2$~mag intervals.  The error bars show the $1 \, \sigma$ uncertainty of
counting statistics, and the open circle represents the place where
incompleteness is becoming important. The curve shows the empirical
function shifted to the most likely distance modulus.  Although this
curve is the ``best fit'' to the data, it is still excluded at the
99.9\% confidence level.}

\figcaption[]  {The open squares show the planetary nebula luminosity
function for PN with isophotal radii between $2\arcmin$ and $4\arcmin$.
The solid circles show a similar dataset, but for objects with 
$r_{\rm iso} > 4\arcmin$.   Both sets of data have been binned into $0.2$~mag 
intervals and have the error bars that represent the $1\, \sigma$ 
uncertainties of counting statistics.  Note that although the brightest
PN is part of the inner sample, the vast majority of bright objects are found
at large galactocentric radii.  In fact, the hypothesis that both sets of
data are drawn from the same underlying distribution is excluded at the
92\% confidence level.}

\figcaption[]  {The circles show the planetary nebula luminosity function
for a complete set of M87 planetaries with $m_{5007} < 27.15$ and isophotal
radii between $2\arcmin$ and $4\arcmin$; the squares show the PNLF for 
intracluster objects scaled to the same area of the galactic survey.  The
$1 \, \sigma$ uncertainties due to Poisson statistics are shown by error 
bars, and the open square and circle denote data past the limit of
completeness.  The dotted curve represents a smooth fit to the intracluster
PNLF, while the solid curve shows the most likely galactic plus intergalactic
luminosity function.  Note that this combined luminosity function fits the
observed PNLF very well.}
 
\figcaption[] {The results of the maximum likelihood analysis for M87.  The
abscissa is the true distance modulus; the ordinate is the relative
probability that the observed PNLF is drawn from a combination of the
intracluster ``field'' PNLF and the empirical model (Ciardullo \etal 1989a)
at the given distance.  In the plot, we have assumed a differential
extinction $E(B$$-$$V) = 0.023$ and a Seaton (1979) reddening law.}

\figcaption[] {The luminosity function of intracluster planetary nebulae
compared to two cluster models.  The solid curve gives the PNLF
expected from an isothermal cluster with core radius $1\pdegree 7$ and
center at $\alpha(1950) = 12^{\rm h} 15^{\rm m}, \ \delta(1950) = +13^\circ$
(Binggeli, Tammann, \& Sandage 1987).  Note that this model is much too
condensed to fit the data.  A better model, displayed by the dotted line,
is one in which the intracluster PN are assumed to have a uniform density
and extend $\sim 4$~Mpc from M87, but even this law is not an adequate
representation of the data.}

\begin{deluxetable}{lllcc}
\tablenum{1}
\tablewidth{0pt}
\tablecaption{Planetary Nebulae Sample}

\tablehead{
\colhead{ID}           & \colhead{$\alpha$ (2000)}      &
\colhead{$\delta$ (2000)}          & \colhead{$m_{5007}$} & 
\colhead{$R_{\rm iso}$}}

\startdata

  1 &      12 30 49.38 & 12 20 58.5 & 25.63 &     2.58 \nl
  2 &      12 30 52.45 & 12 21 13.5 & 26.11 &     2.37 \nl
  3 &      12 30 45.49 & 12 22 01.5 & 26.16 &     1.92 \nl
  4 &      12 30 46.08 & 12 19 37.9 & 26.05 &     4.28 \nl
  5 &      12 30 43.87 & 12 20 39.2 & 26.58 &     3.58 \nl
  6 &      12 30 49.69 & 12 22 04.3 & 26.62 &     1.42 \nl
  7 &      12 30 51.55 & 12 19 40.7 & 26.35 &     3.89 \nl
  8 &      12 30 47.58 & 12 21 25.7 & 26.48 &     2.22 \nl
  9 &      12 30 52.85 & 12 25 10.0 & 26.45 &     2.17 \nl
 10 &      12 30 47.98 & 12 21 12.1 & 27.03 &     2.43 \nl
 11 &      12 30 50.47 & 12 20 06.3 & 26.60 &     3.45 \nl
 12 &      12 30 46.96 & 12 25 18.9 & 26.64 &     1.95 \nl
 13 &      12 30 50.20 & 12 25 09.0 & 26.41 &     1.81 \nl
 14 &      12 30 47.03 & 12 25 05.2 & 27.12 &     1.73 \nl
 15 &      12 30 56.52 & 12 24 15.0 & 26.62 &     2.35 \nl
 16 &      12 30 54.75 & 12 20 52.0 & 26.98 &     2.92 \nl
 17 &      12 30 46.94 & 12 24 35.9 & 26.92 &     1.28 \nl
 18 &      12 30 53.18 & 12 23 54.0 & 26.72 &     1.19 \nl
 19 &      12 30 54.28 & 12 20 02.1 & 27.12 &     3.63 \nl
 20 &      12 30 44.42 & 12 23 47.3 & 26.92 &     1.32 \nl
 21 &      12 30 51.48 & 12 19 53.8 & 26.58 &     3.66 \nl
 22 &      12 30 56.40 & 12 21 25.8 & 26.94 &     2.73 \nl
 23 &      12 30 54.51 & 12 22 14.6 & 26.97 &     1.80 \nl
 24 &      12 30 54.51 & 12 22 14.6 & 26.97 &     1.80 \nl
 25 &      12 30 43.85 & 12 21 23.8 & 27.10 &     2.86 \nl
 26 &      12 30 53.27 & 12 21 19.2 & 26.94 &     2.35 \nl
 27 &      12 30 54.69 & 12 24 06.3 & 26.43 &     1.72 \nl
 28 &      12 30 56.37 & 12 19 05.9 & 26.95 &     4.69 \nl
 29 &      12 30 42.99 & 12 24 29.5 & 26.72 &     1.93 \nl
 31 &      12 30 47.74 & 12 24 43.1 & 26.53 &     1.32 \nl
 32 &      12 30 50.79 & 12 21 16.6 & 27.09 &     2.25 \nl
 33 &      12 30 48.22 & 12 21 44.5 & 27.24 &     1.83 \nl
 34 &      12 30 50.25 & 12 21 02.2 & 26.91 &     2.49 \nl
 36 &      12 30 55.20 & 12 21 31.0 & 27.38 &     2.44 \nl
 37 &      12 30 46.64 & 12 24 30.3 & 26.93 &     1.23 \nl
 38 &      12 30 52.33 & 12 20 32.2 & 27.21 &     3.03 \nl
 39 &      12 30 46.55 & 12 21 56.3 & 26.98 &     1.82 \nl
 40 &      12 30 55.02 & 12 23 44.7 & 27.05 &     1.66 \nl
 41 &      12 30 54.68 & 12 22 14.1 & 27.16 &     1.84 \nl 
 42 &      12 30 50.84 & 12 24 45.7 & 26.70 &     1.47 \nl
 43 &      12 30 50.01 & 12 21 09.1 & 26.72 &     2.38 \nl
 44 &      12 30 42.81 & 12 23 51.1 & 26.27 &     1.78 \nl
 46 &      12 30 45.17 & 12 24 58.7 & 27.37 &     1.83 \nl
 47 &      12 30 52.36 & 12 20 08.0 & 27.58 &     3.43 \nl
 48 &      12 30 44.25 & 12 24 59.8 & 27.35 &     1.99 \nl
 56 &      12 30 46.67 & 12 23 18.5 & 25.92 &     0.69 \nl
 57 &      12 30 51.41 & 12 22 59.0 & 25.94 &     0.72 \nl
 58 &      12 31 07.02 & 12 20 02.9 & 25.97 &     5.97 \nl
 59 &      12 30 30.88 & 12 21 52.1 & 26.00 &     6.18 \nl
 60 &      12 30 38.96 & 12 28 13.8 & 26.01 &     5.42 \nl
 61 &      12 30 35.04 & 12 24 14.1 & 26.10 &     4.31 \nl
 62 &      12 30 47.36 & 12 18 40.5 & 26.12 &     5.13 \nl
 63 &      12 30 44.84 & 12 23 29.7 & 26.13 &     1.18 \nl
 64 &      12 30 42.14 & 12 29 13.9 & 26.21 &     6.05 \nl
 65 &      12 30 43.21 & 12 24 15.1 & 26.21 &     1.77 \nl
 66 &      12 30 36.87 & 12 26 17.6 & 26.22 &     4.39 \nl
 67 &      12 30 43.48 & 12 23 31.8 & 26.22 &     1.58 \nl
 68 &      12 30 34.76 & 12 20 21.2 & 26.23 &     5.97 \nl
 69 &      12 30 57.55 & 12 19 23.2 & 26.25 &     4.55 \nl
 70 &      12 30 57.91 & 12 21 42.0 & 26.25 &     2.89 \nl
 71 &      12 30 18.61 & 12 22 03.2 & 26.26 &     9.81 \nl
 72 &      12 30 36.12 & 12 27 26.6 & 26.27 &     5.27 \nl
 73 &      12 30 36.24 & 12 26 57.5 & 26.29 &     4.92 \nl
 74 &      12 30 43.03 & 12 22 45.7 & 26.29 &     1.97 \nl
 75 &      12 30 52.01 & 12 23 30.8 & 26.29 &     0.72 \nl
 76 &      12 31 08.64 & 12 21 22.5 & 26.30 &     5.98 \nl
 77 &      12 31 02.92 & 12 30 02.9 & 26.30 &     8.71 \nl
 78 &      12 30 46.10 & 12 27 52.7 & 26.31 &     4.56 \nl
 79 &      12 30 27.42 & 12 22 16.7 & 26.31 &     7.05 \nl
 80 &      12 30 55.50 & 12 15 44.4 & 26.31 &     7.96 \nl
 81 &      12 30 53.52 & 12 22 06.8 & 26.32 &     1.71 \nl
 82 &      12 31 01.77 & 12 21 08.1 & 26.32 &     4.17 \nl
 83 &      12 31 11.24 & 12 21 17.5 & 26.32 &     6.74 \nl
 84 &      12 30 48.24 & 12 18 33.8 & 26.39 &     5.18 \nl
 85 &      12 30 51.24 & 12 27 53.5 & 26.39 &     4.79 \nl
 86 &      12 30 58.09 & 12 23 32.6 & 26.40 &     2.57 \nl
 87 &      12 30 37.15 & 12 19 08.1 & 26.41 &     6.36 \nl
 88 &      12 30 42.44 & 12 23 00.8 & 26.42 &     2.02 \nl
 89 &      12 30 39.67 & 12 24 25.9 & 26.44 &     2.83 \nl
 90 &      12 31 02.62 & 12 25 56.1 & 26.46 &     5.25 \nl
 91 &      12 30 31.32 & 12 30 04.7 & 26.46 &     8.05 \nl
 92 &      12 31 06.64 & 12 24 41.6 & 26.47 &     5.73 \nl
 93 &      12 30 52.29 & 12 23 37.5 & 26.49 &     0.83 \nl
 94 &      12 31 00.83 & 12 27 01.1 & 26.50 &     5.61 \nl
 95 &      12 30 46.76 & 12 20 40.3 & 26.50 &     3.11 \nl
 96 &      12 30 45.61 & 12 23 35.9 & 26.51 &     0.96 \nl
 97 &      12 30 38.43 & 12 23 03.1 & 26.51 &     3.34 \nl
 98 &      12 30 35.41 & 12 24 45.5 & 26.52 &     4.24 \nl
 99 &      12 30 35.21 & 12 26 16.6 & 26.52 &     4.78 \nl
100 &      12 31 07.63 & 12 17 31.9 & 26.52 &     7.60 \nl
101 &      12 31 03.73 & 12 19 50.2 & 26.52 &     5.31 \nl
102 &      12 30 20.30 & 12 26 11.8 & 26.53 &     8.86 \nl
103 &      12 30 54.65 & 12 26 07.8 & 26.53 &     3.46 \nl
104 &      12 31 07.42 & 12 24 30.1 & 26.53 &     5.89 \nl
105 &      12 30 54.57 & 12 23 38.5 & 26.55 &     1.49 \nl
106 &      12 31 02.52 & 12 27 11.3 & 26.55 &     6.14 \nl
107 &      12 30 45.43 & 12 26 14.1 & 26.55 &     2.94 \nl
108 &      12 30 41.77 & 12 23 38.6 & 26.56 &     2.09 \nl
109 &      12 30 40.50 & 12 25 20.8 & 26.56 &     2.99 \nl
110 &      12 30 44.44 & 12 23 06.4 & 26.56 &     1.39 \nl
111 &      12 31 04.31 & 12 27 01.9 & 26.57 &     6.47 \nl
112 &      12 30 52.28 & 12 17 45.2 & 26.59 &     5.87 \nl
113 &      12 30 41.97 & 12 24 36.0 & 26.59 &     2.23 \nl
114 &      12 30 37.61 & 12 23 02.9 & 26.59 &     3.63 \nl
115 &      12 30 37.73 & 12 26 43.4 & 26.59 &     4.47 \nl
116 &      12 31 05.07 & 12 19 55.7 & 26.60 &     5.56 \nl
117 &      12 30 48.05 & 12 24 10.3 & 26.60 &     0.78 \nl
118 &      12 30 48.14 & 12 24 52.1 & 26.60 &     1.46 \nl
119 &      12 31 08.74 & 12 23 41.7 & 26.61 &     6.05 \nl
120 &      12 30 37.03 & 12 19 08.7 & 26.62 &     6.37 \nl
121 &      12 30 42.62 & 12 24 31.9 & 26.63 &     2.04 \nl
122 &      12 30 47.41 & 12 22 39.6 & 26.63 &     0.97 \nl
123 &      12 30 30.12 & 12 23 47.5 & 26.63 &     5.84 \nl
124 &      12 30 28.42 & 12 28 54.7 & 26.64 &     7.78 \nl
125 &      12 30 52.66 & 12 23 16.1 & 26.64 &     0.90 \nl
126 &      12 30 52.47 & 12 25 38.3 & 26.64 &     2.59 \nl
127 &      12 30 41.00 & 12 21 33.8 & 26.64 &     3.42 \nl
128 &      12 30 59.99 & 12 21 23.0 & 26.64 &     3.59 \nl
129 &      12 30 33.62 & 12 26 42.3 & 26.65 &     5.37 \nl
130 &      12 30 45.34 & 12 25 31.3 & 26.65 &     2.28 \nl
131 &      12 31 02.98 & 12 21 48.4 & 26.66 &     4.27 \nl
132 &      12 30 51.70 & 12 22 09.8 & 26.66 &     1.42 \nl
133 &      12 30 19.85 & 12 25 06.5 & 26.66 &     8.90 \nl
134 &      12 31 10.32 & 12 26 19.2 & 26.66 &     7.62 \nl
135 &      12 30 46.45 & 12 22 49.7 & 26.68 &     1.01 \nl
136 &      12 30 50.43 & 12 26 06.7 & 26.69 &     2.85 \nl
137 &      12 30 33.19 & 12 24 01.4 & 26.70 &     4.88 \nl
138 &      12 31 11.63 & 12 16 25.1 & 26.70 &     9.13 \nl
139 &      12 30 49.25 & 12 18 46.7 & 26.70 &     4.89 \nl
140 &      12 30 44.78 & 12 22 57.2 & 26.71 &     1.37 \nl
141 &      12 30 47.65 & 12 24 36.3 & 26.71 &     1.22 \nl
142 &      12 31 06.39 & 12 23 54.7 & 26.71 &     5.39 \nl
143 &      12 30 52.70 & 12 24 11.2 & 26.73 &     1.25 \nl
144 &      12 30 53.74 & 12 26 53.7 & 26.73 &     4.08 \nl
145 &      12 31 07.38 & 12 21 41.0 & 26.73 &     5.56 \nl
146 &      12 30 53.87 & 12 24 02.0 & 26.73 &     1.46 \nl
147 &      12 30 40.06 & 12 25 25.1 & 26.74 &     3.14 \nl
148 &      12 30 44.04 & 12 28 30.5 & 26.74 &     5.24 \nl
149 &      12 30 48.66 & 12 22 34.6 & 26.74 &     0.91 \nl
150 &      12 30 37.55 & 12 29 54.5 & 26.74 &     7.06 \nl
151 &      12 30 52.92 & 12 23 50.2 & 26.74 &     1.08 \nl
152 &      12 30 21.78 & 12 16 39.3 & 26.75 &     2.00 \nl
153 &      12 30 59.73 & 12 19 16.4 & 26.75 &     4.95 \nl
154 &      12 30 26.18 & 12 23 59.8 & 26.75 &     7.02 \nl
155 &      12 30 46.85 & 12 20 43.2 & 26.75 &     3.05 \nl
156 &      12 30 41.93 & 12 26 48.6 & 26.76 &     3.82 \nl
157 &      12 30 51.15 & 12 22 26.7 & 26.76 &     1.11 \nl
158 &      12 30 52.28 & 12 20 59.9 & 26.77 &     2.58 \nl
159 &      12 30 34.65 & 12 25 53.8 & 26.77 &     4.76 \nl
160 &      12 30 57.25 & 12 22 43.5 & 26.77 &     2.30 \nl
161 &      12 30 35.64 & 12 30 22.7 & 26.77 &     7.69 \nl
162 &      12 30 49.20 & 12 21 51.6 & 26.79 &     1.65 \nl
163 &      12 30 36.91 & 12 24 07.3 & 26.79 &     3.70 \nl
164 &      12 31 04.55 & 12 24 58.2 & 26.80 &     5.24 \nl
165 &      12 30 53.75 & 12 28 04.2 & 26.80 &     5.26 \nl
166 &      12 30 40.16 & 12 27 08.9 & 26.80 &     4.35 \nl
167 &      12 30 45.69 & 12 29 04.7 & 26.81 &     5.79 \nl
168 &      12 30 53.15 & 12 20 55.9 & 26.81 &     2.70 \nl
169 &      12 30 40.19 & 12 27 15.1 & 26.81 &     4.43 \nl
170 &      12 31 09.26 & 12 27 08.7 & 26.81 &     7.84 \nl
171 &      12 30 41.48 & 12 23 03.5 & 26.82 &     2.31 \nl
172 &      12 30 46.38 & 12 23 13.5 & 26.83 &     0.79 \nl
173 &      12 30 45.89 & 12 24 38.4 & 26.83 &     1.44 \nl
174 &      12 31 17.38 & 12 22 07.6 & 26.83 &     8.52 \nl
175 &      12 30 42.72 & 12 25 29.2 & 26.83 &     2.62 \nl
176 &      12 30 51.83 & 12 23 48.9 & 26.84 &     0.78 \nl
177 &      12 30 40.20 & 12 25 30.1 & 26.83 &     3.16 \nl
178 &      12 30 42.33 & 12 20 54.4 & 26.84 &     3.67 \nl
179 &      12 30 47.58 & 12 25 54.1 & 26.84 &     2.53 \nl
180 &      12 30 46.84 & 12 24 09.8 & 26.85 &     0.93 \nl
181 &      12 30 40.92 & 12 23 18.8 & 26.85 &     2.41 \nl
182 &      12 30 27.24 & 12 20 54.5 & 26.86 &     7.72 \nl
183 &      12 30 53.62 & 12 27 49.9 & 26.86 &     5.00 \nl
184 &      12 31 00.71 & 12 29 14.1 & 26.86 &     7.54 \nl
185 &      12 31 08.95 & 12 23 17.3 & 26.86 &     6.04 \nl
186 &      12 30 59.94 & 12 22 41.5 & 26.87 &     3.16 \nl
187 &      12 30 45.22 & 12 26 28.2 & 26.89 &     3.18 \nl
188 &      12 30 33.98 & 12 28 50.9 & 26.89 &     6.66 \nl
189 &      12 30 47.07 & 12 24 09.9 & 26.89 &     0.89 \nl
190 &      12 30 47.71 & 12 24 12.7 & 26.89 &     0.85 \nl
191 &      12 30 51.46 & 12 21 37.0 & 26.89 &     1.92 \nl
192 &      12 30 57.89 & 12 24 50.6 & 26.89 &     3.15 \nl
193 &      12 30 41.24 & 12 24 10.9 & 26.89 &     2.30 \nl
194 &      12 30 45.74 & 12 24 27.8 & 26.90 &     1.33 \nl
195 &      12 30 39.47 & 12 25 50.6 & 26.91 &     3.53 \nl
196 &      12 30 40.12 & 12 28 16.6 & 26.91 &     5.32 \nl
197 &      12 30 52.23 & 12 24 29.8 & 26.92 &     1.42 \nl
198 &      12 30 57.80 & 12 27 25.6 & 26.92 &     5.31 \nl
199 &      12 30 45.08 & 12 22 56.1 & 26.92 &     1.29 \nl
200 &      12 30 47.84 & 12 25 51.4 & 26.93 &     2.48 \nl
201 &      12 30 42.52 & 12 18 29.8 & 26.93 &     5.92 \nl
202 &      12 30 57.62 & 12 25 43.7 & 26.93 &     3.79 \nl
203 &      12 30 42.37 & 12 25 37.9 & 26.94 &     2.79 \nl
204 &      12 30 53.75 & 12 17 19.8 & 26.94 &     6.31 \nl
205 &      12 30 52.60 & 12 27 01.7 & 26.95 &     4.05 \nl
206 &      12 30 23.10 & 12 26 34.7 & 26.95 &     8.12 \nl
207 &      12 30 45.66 & 12 25 59.5 & 26.96 &     2.69 \nl
208 &      12 30 38.43 & 12 22 25.2 & 26.96 &     3.63 \nl
209 &      12 30 43.14 & 12 18 06.4 & 26.98 &     6.22 \nl
210 &      12 31 00.86 & 12 23 02.5 & 26.98 &     3.46 \nl
211 &      12 31 04.02 & 12 21 54.0 & 26.98 &     4.55 \nl
212 &      12 30 46.47 & 12 27 44.9 & 26.98 &     4.42 \nl
213 &      12 31 09.30 & 12 16 59.2 & 26.99 &     8.30 \nl
214 &      12 30 42.36 & 12 15 32.7 & 26.99 &     8.91 \nl
215 &      12 30 51.93 & 12 18 02.4 & 27.00 &     5.57 \nl
216 &      12 30 18.53 & 12 24 13.1 & 27.00 &     9.35 \nl
217 &      12 30 47.57 & 12 21 40.5 & 27.00 &     1.96 \nl
218 &      12 30 42.16 & 12 20 28.6 & 27.00 &     4.11 \nl
219 &      12 30 43.92 & 12 19 30.3 & 27.00 &     4.71 \nl
220 &      12 31 08.61 & 12 27 39.8 & 27.00 &     8.03 \nl
221 &      12 30 37.94 & 12 25 43.3 & 27.01 &     3.83 \nl
222 &      12 30 54.42 & 12 24 48.5 & 27.00 &     2.17 \nl
223 &      12 30 52.87 & 12 21 14.8 & 27.01 &     2.38 \nl
224 &      12 30 52.64 & 12 25 30.6 & 27.01 &     2.49 \nl
225 &      12 30 32.08 & 12 20 56.9 & 27.01 &     6.31 \nl
226 &      12 30 44.07 & 12 23 54.1 & 27.02 &     1.44 \nl
227 &      12 30 51.42 & 12 26 54.0 & 27.03 &     3.78 \nl
228 &      12 31 18.45 & 12 18 13.9 & 27.03 &     9.64 \nl
229 &      12 30 27.84 & 12 17 02.2 & 27.03 &     0.23 \nl
230 &      12 30 40.40 & 12 20 52.0 & 27.04 &     4.17 \nl
231 &      12 30 50.66 & 12 25 06.7 & 27.05 &     1.81 \nl
232 &      12 30 44.64 & 12 21 49.5 & 27.05 &     2.29 \nl
233 &      12 30 42.22 & 12 22 16.2 & 27.05 &     2.53 \nl
234 &      12 31 00.21 & 12 25 15.8 & 27.05 &     4.16 \nl
235 &      12 30 32.33 & 12 24 46.1 & 27.05 &     5.15 \nl
236 &      12 30 53.71 & 12 21 37.4 & 27.05 &     2.13 \nl
237 &      12 30 31.93 & 12 18 19.3 & 27.06 &     8.23 \nl
238 &      12 30 44.55 & 12 26 39.1 & 27.07 &     3.40 \nl
239 &      12 30 23.23 & 12 20 03.9 & 27.07 &     9.33 \nl
240 &      12 30 57.69 & 12 28 44.6 & 27.08 &     6.53 \nl
241 &      12 30 32.47 & 12 21 46.2 & 27.08 &     5.75 \nl
242 &      12 30 50.90 & 12 20 57.3 & 27.08 &     2.58 \nl
243 &      12 30 56.88 & 12 21 18.6 & 27.09 &     2.91 \nl
244 &      12 30 29.30 & 12 26 12.7 & 27.09 &     6.30 \nl
245 &      12 30 44.09 & 12 23 30.0 & 27.09 &     1.40 \nl
246 &      12 30 27.31 & 12 21 33.2 & 27.10 &     7.38 \nl
247 &      12 30 56.39 & 12 23 11.0 & 27.10 &     1.98 \nl
248 &      12 30 41.04 & 12 25 49.4 & 27.10 &     3.18 \nl
249 &      12 30 46.16 & 12 22 36.2 & 27.11 &     1.26 \nl
250 &      12 30 52.82 & 12 27 05.2 & 27.11 &     4.14 \nl
251 &      12 30 41.77 & 12 23 36.3 & 27.11 &     2.09 \nl
252 &      12 30 47.16 & 12 19 13.1 & 27.11 &     4.59 \nl
253 &      12 30 38.27 & 12 24 58.8 & 27.12 &     3.42 \nl
254 &      12 31 02.85 & 12 19 38.8 & 27.12 &     5.24 \nl
255 &      12 30 44.92 & 12 25 56.1 & 27.12 &     2.70 \nl
256 &      12 30 57.73 & 12 24 05.4 & 27.13 &     2.65 \nl
257 &      12 30 47.52 & 12 20 07.9 & 27.13 &     3.60 \nl
258 &      12 30 39.93 & 12 21 39.4 & 27.13 &     3.66 \nl
259 &      12 31 10.83 & 12 20 34.8 & 27.14 &     6.78 \nl
260 &      12 30 57.80 & 12 25 42.6 & 27.14 &     3.82 \nl
261 &      12 30 44.10 & 12 22 09.8 & 27.14 &     2.11 \nl
262 &      12 30 53.98 & 12 21 39.8 & 27.15 &     2.14 \nl
263 &      12 30 44.74 & 12 21 41.9 & 27.15 &     2.39 \nl
264 &      12 30 32.45 & 12 19 04.3 & 27.16 &     7.51 \nl
265 &      12 30 39.26 & 12 24 59.7 & 27.16 &     3.14 \nl
266 &      12 30 56.74 & 12 21 25.4 & 27.17 &     2.80 \nl
267 &      12 30 42.39 & 12 25 41.5 & 27.17 &     2.83 \nl
268 &      12 30 58.42 & 12 19 55.4 & 27.17 &     4.23 \nl
269 &      12 31 05.60 & 12 23 59.9 & 27.17 &     5.17 \nl
270 &      12 30 57.87 & 12 24 42.5 & 27.18 &     3.05 \nl
271 &      12 30 54.61 & 12 15 43.5 & 27.18 &     7.96 \nl
272 &      12 30 57.72 & 12 22 15.4 & 27.18 &     2.58 \nl
273 &      12 30 43.29 & 12 25 55.8 & 27.18 &     2.88 \nl
274 &      12 30 39.57 & 12 23 35.1 & 27.19 &     2.82 \nl
275 &      12 30 55.00 & 12 18 24.3 & 27.19 &     5.26 \nl
276 &      12 30 45.78 & 12 22 48.5 & 27.20 &     1.19 \nl
277 &      12 30 50.41 & 12 19 43.2 & 27.20 &     3.86 \nl
278 &      12 30 29.59 & 12 28 01.4 & 27.21 &     7.00 \nl
279 &      12 30 35.80 & 12 18 51.7 & 27.21 &     6.89 \nl
280 &      12 30 54.42 & 12 16 28.2 & 27.21 &     7.20 \nl
281 &      12 31 10.11 & 12 17 48.9 & 27.22 &     7.90 \nl
282 &      12 31 08.18 & 12 19 24.5 & 27.22 &     6.56 \nl
283 &      12 30 25.64 & 12 19 43.5 & 27.22 &     8.85 \nl
284 &      12 30 58.62 & 12 25 07.3 & 27.22 &     3.58 \nl
285 &      12 30 55.38 & 12 26 57.1 & 27.23 &     4.41 \nl
286 &      12 30 53.86 & 12 16 21.3 & 27.23 &     7.31 \nl
287 &      12 30 48.69 & 12 21 51.4 & 27.23 &     1.68 \nl
288 &      12 30 52.85 & 12 26 04.6 & 27.23 &     3.09 \nl
289 &      12 30 41.12 & 12 24 05.4 & 27.24 &     2.32 \nl
290 &      12 30 45.93 & 12 21 20.7 & 27.24 &     2.53 \nl
291 &      12 31 08.37 & 12 27 06.6 & 27.25 &     7.58 \nl
292 &      12 30 55.14 & 12 18 45.3 & 27.25 &     4.92 \nl
293 &      12 30 21.35 & 12 18 03.9 & 27.25 &     1.10 \nl
294 &      12 30 41.50 & 12 17 33.7 & 27.25 &     7.01 \nl
295 &      12 30 54.53 & 12 21 09.7 & 27.26 &     2.63 \nl
296 &      12 30 59.26 & 12 19 28.1 & 27.26 &     4.72 \nl
297 &      12 31 00.21 & 12 23 21.9 & 27.27 &     3.27 \nl
298 &      12 30 39.13 & 12 28 20.9 & 27.27 &     5.50 \nl
299 &      12 30 51.77 & 12 26 29.7 & 27.28 &     3.38 \nl
300 &      12 30 44.92 & 12 19 52.0 & 27.28 &     4.20 \nl
301 &      12 30 45.89 & 12 24 25.5 & 27.29 &     1.28 \nl
302 &      12 30 58.85 & 12 23 25.9 & 27.30 &     2.81 \nl
303 &      12 30 39.75 & 12 22 43.5 & 27.31 &     3.03 \nl
304 &      12 30 42.79 & 12 18 41.3 & 27.34 &     5.69 \nl
305 &      12 30 37.67 & 12 21 37.0 & 27.35 &     4.35 \nl
306 &      12 31 00.82 & 12 17 04.0 & 27.36 &     6.99 \nl
307 &      12 30 52.95 & 12 20 05.3 & 27.38 &     3.50 \nl
308 &      12 30 55.80 & 12 21 16.2 & 27.38 &     2.74 \nl
309 &      12 30 25.43 & 12 27 11.9 & 27.39 &     7.67 \nl
310 &      12 30 50.33 & 12 18 59.3 & 27.39 &     4.62 \nl
311 &      12 30 40.34 & 12 23 43.3 & 27.40 &     2.55 \nl
312 &      12 30 43.97 & 12 20 35.3 & 27.40 &     3.63 \nl
313 &      12 30 41.84 & 12 22 53.6 & 27.41 &     2.27 \nl
314 &      12 30 55.58 & 12 20 27.2 & 27.42 &     3.38 \nl
315 &      12 30 29.80 & 12 22 07.8 & 27.43 &     6.39 \nl
316 &      12 31 09.30 & 12 18 21.9 & 27.43 &     7.39 \nl
317 &      12 30 40.75 & 12 23 01.6 & 27.44 &     2.56 \nl
318 &      12 30 59.65 & 12 19 31.2 & 27.45 &     4.74 \nl
319 &      12 30 59.62 & 12 22 06.1 & 27.45 &     3.19 \nl
320 &      12 30 58.68 & 12 25 27.3 & 27.46 &     3.86 \nl
321 &      12 30 42.12 & 12 18 46.9 & 27.47 &     5.71 \nl
322 &      12 30 48.80 & 12 21 32.5 & 27.52 &     2.00 \nl
323 &      12 30 41.55 & 12 23 20.0 & 27.52 &     2.20 \nl
324 &      12 30 47.65 & 12 25 45.8 & 27.52 &     2.38 \nl
325 &      12 30 56.30 & 12 23 33.8 & 27.52 &     2.00 \nl
326 &      12 31 03.60 & 12 25 43.1 & 27.55 &     5.38 \nl
327 &      12 30 43.40 & 12 21 26.0 & 27.55 &     2.93 \nl
328 &      12 31 00.47 & 12 24 46.9 & 27.56 &     3.94 \nl
329 &      12 30 53.33 & 12 28 14.6 & 27.57 &     5.38 \nl
330 &      12 31 03.55 & 12 18 56.3 & 27.59 &     5.85 \nl
331 &      12 30 40.20 & 12 19 22.9 & 27.60 &     5.49 \nl
332 &      12 31 01.50 & 12 21 36.4 & 27.61 &     3.91 \nl
333 &      12 30 46.66 & 12 21 39.1 & 27.61 &     2.10 \nl
334 &      12 30 55.99 & 12 20 19.2 & 27.61 &     3.55 \nl
335 &      12 30 26.48 & 12 20 13.9 & 27.61 &     8.31 \nl
336 &      12 30 39.66 & 12 23 58.4 & 27.64 &     2.77 \nl
337 &      12 30 33.19 & 12 20 55.5 & 27.64 &     6.02 \nl
338 &      12 30 42.09 & 12 24 20.2 & 27.65 &     2.10 \nl
339 &      12 30 48.74 & 12 22 02.5 & 27.86 &     1.48 \nl

\enddata
\end{deluxetable}

\begin{deluxetable}{lccc}
\tablenum{2}
\tablewidth{0pt}
\tablecaption{Planetary Nebulae Associated with other galaxies}

\tablehead{
\colhead{ID}           & \colhead{$\alpha$ (2000)}      &
\colhead{$\delta$ (2000)}          & \colhead{$m_{5007}$}}

\startdata

NGC 4478 - 1 & 12 30 24.48 &  12 20 01.3  & 26.18 \nl
NGC 4478 - 2 & 12 30 18.52 &  12 19 21.9  & 26.37 \nl
NGC 4478 - 3 & 12 30 18.45 &  12 19 24.9  & 26.95 \nl
NGC 4478 - 4 & 12 30 18.60 &  12 19 28.5  & 26.79 \nl
NGC 4478 - 5 & 12 30 18.48 &  12 19 30.9  & 26.509 \nl
NGC 4478 - 6 & 12 30 18.57 &  12 19 58.4  & 27.04 \nl
NGC 4478 - 7 & 12 30 19.10 &  12 19 17.3  & 27.65 \nl
IC 3443 - 1  & 12 31 13.20 &  12 19 55.2  & 26.73 \nl
Anonymous - 1  & 12 30 27.29 &  12 16 54.9  & 26.69 \nl

\enddata
\end{deluxetable}

\begin{deluxetable}{lcc}
\tablenum{3}
\tablewidth{0pt}
\tablecaption{PN Photometric Error Versus Magnitude}

\tablehead{ \colhead{Magnitude} & \colhead{Mean 1 $\sigma$ error}
& \colhead{Number}}

\startdata
25.6 &     0.036 &     1 \nl
26.0 &     0.050 &     4 \nl
26.2 &     0.056 &    12 \nl
26.4 &     0.061 &    18 \nl
26.6 &     0.080 &    37 \nl
26.8 &     0.093 &    40 \nl
27.0 &     0.110 &    52 \nl
27.2 &     0.125 &    48 \nl
27.4 &     0.142 &    21 \nl
27.6 &     0.171 &    14 \nl
\enddata
\end{deluxetable}

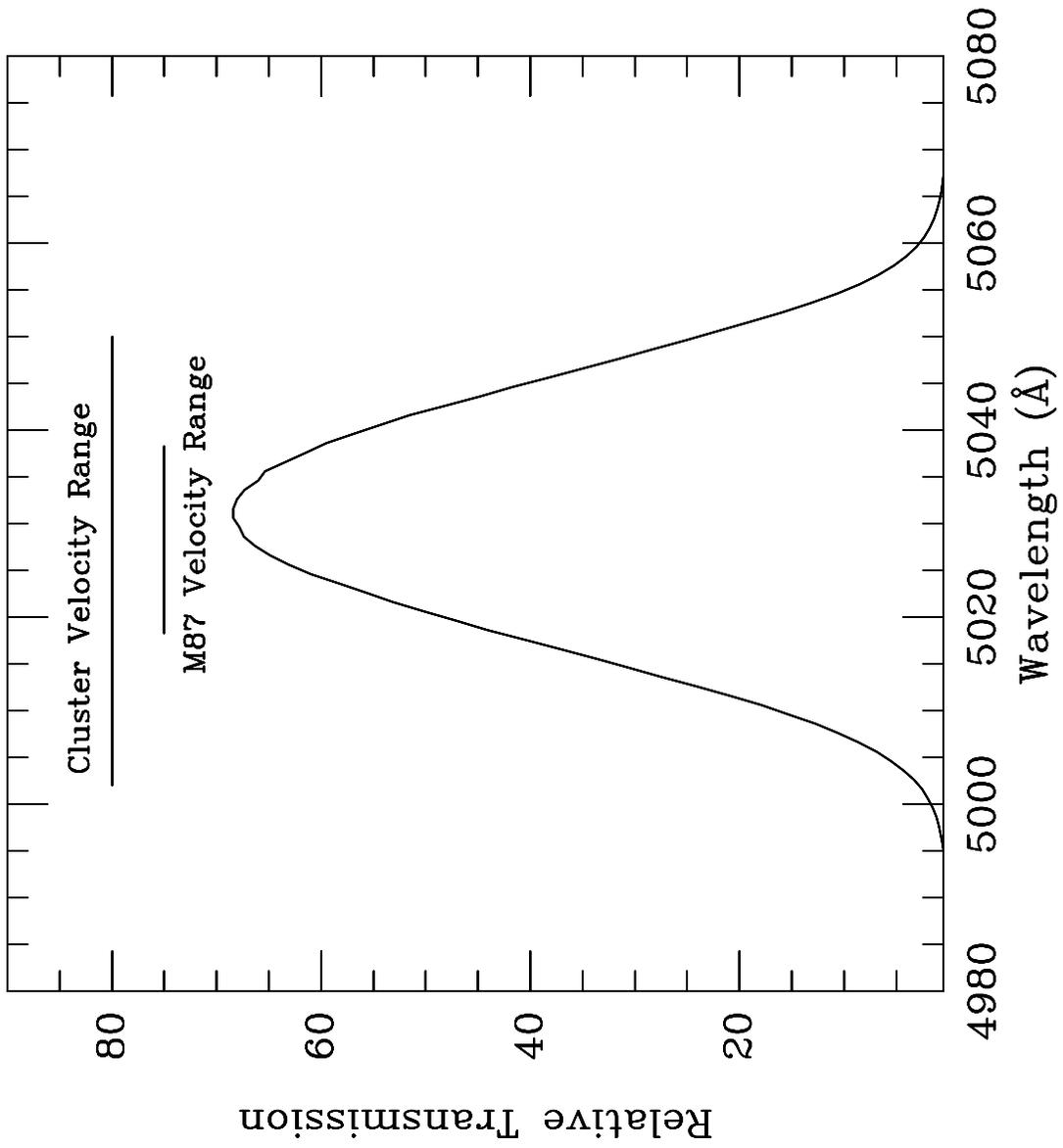
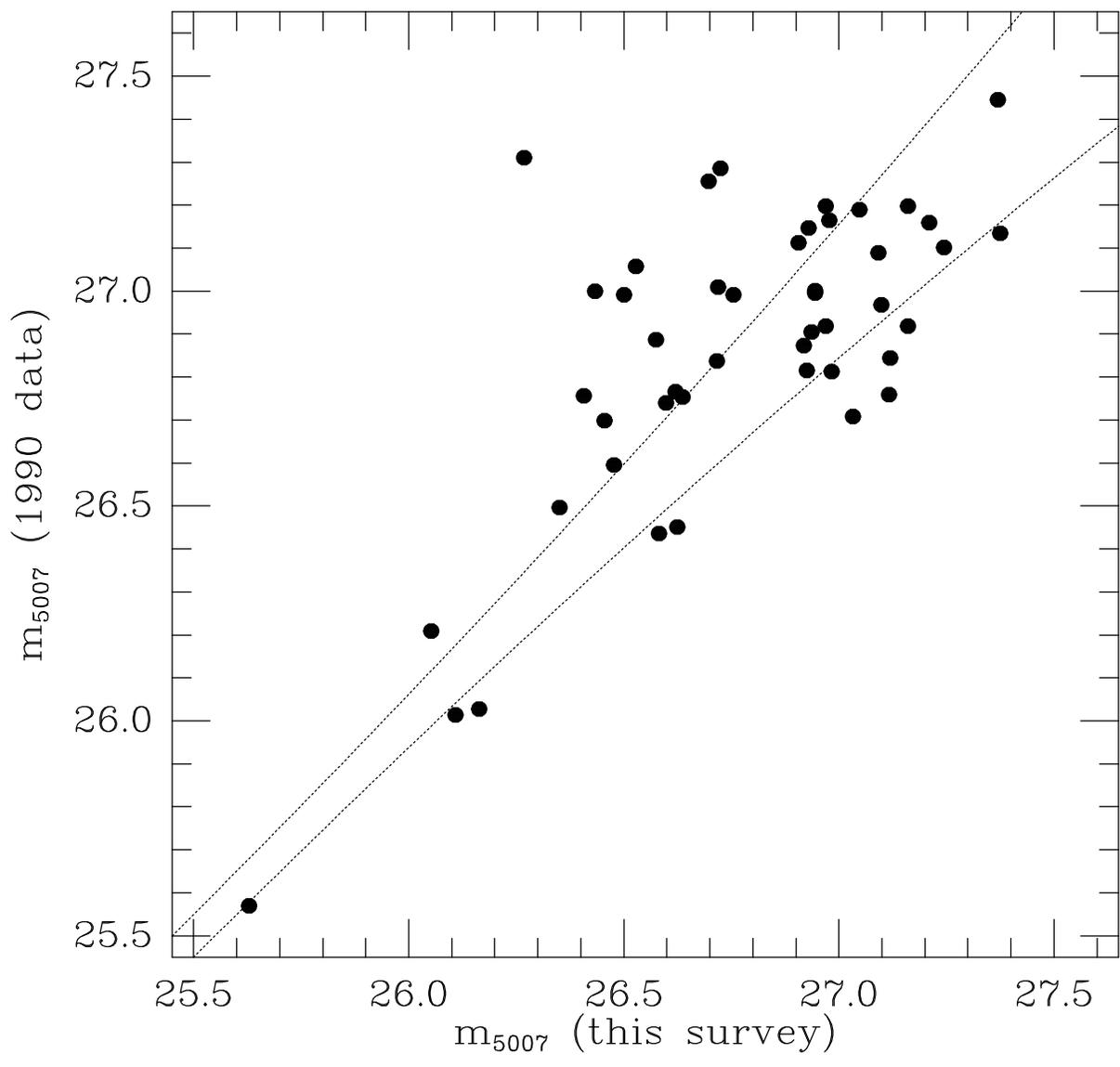
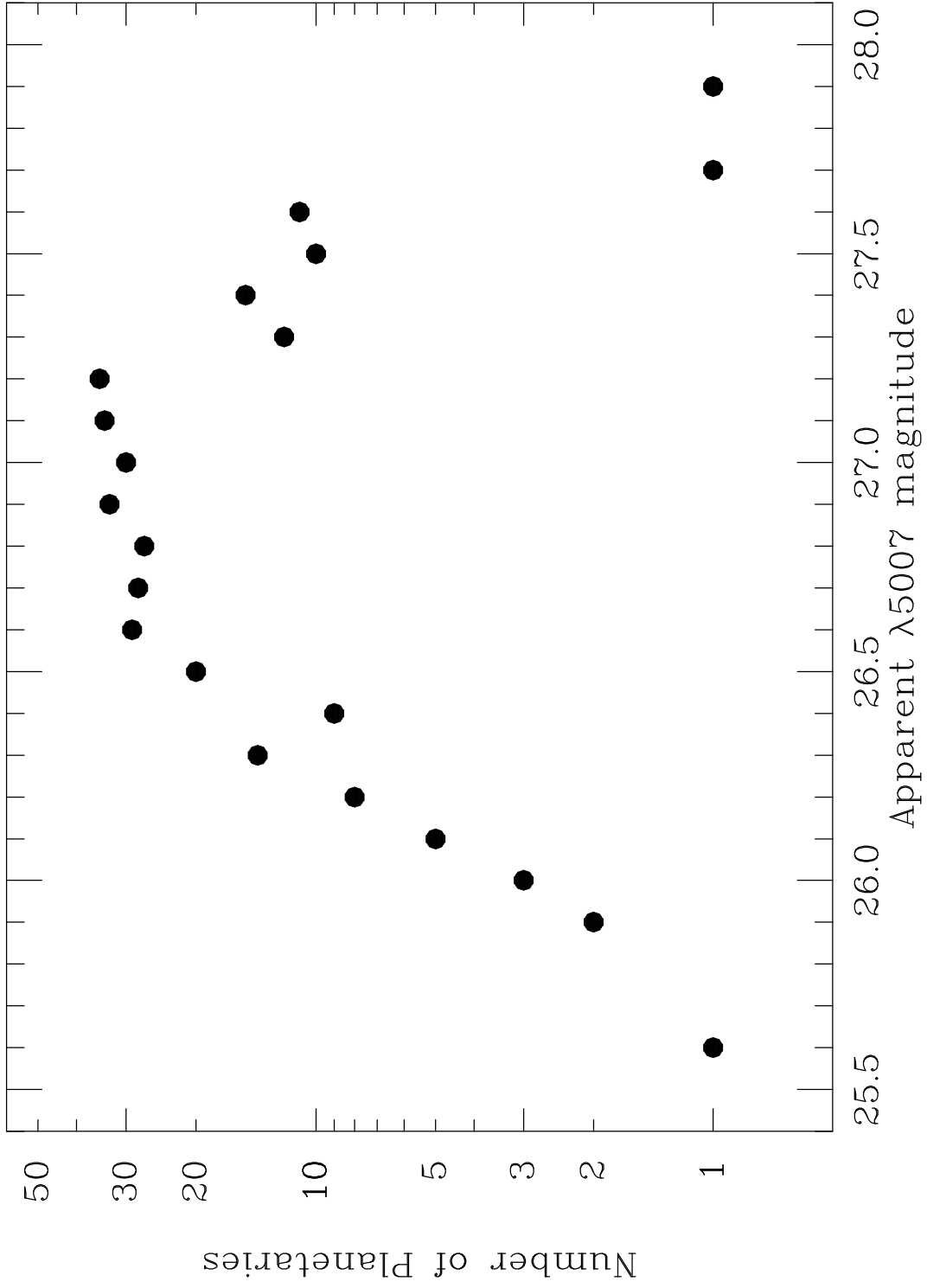
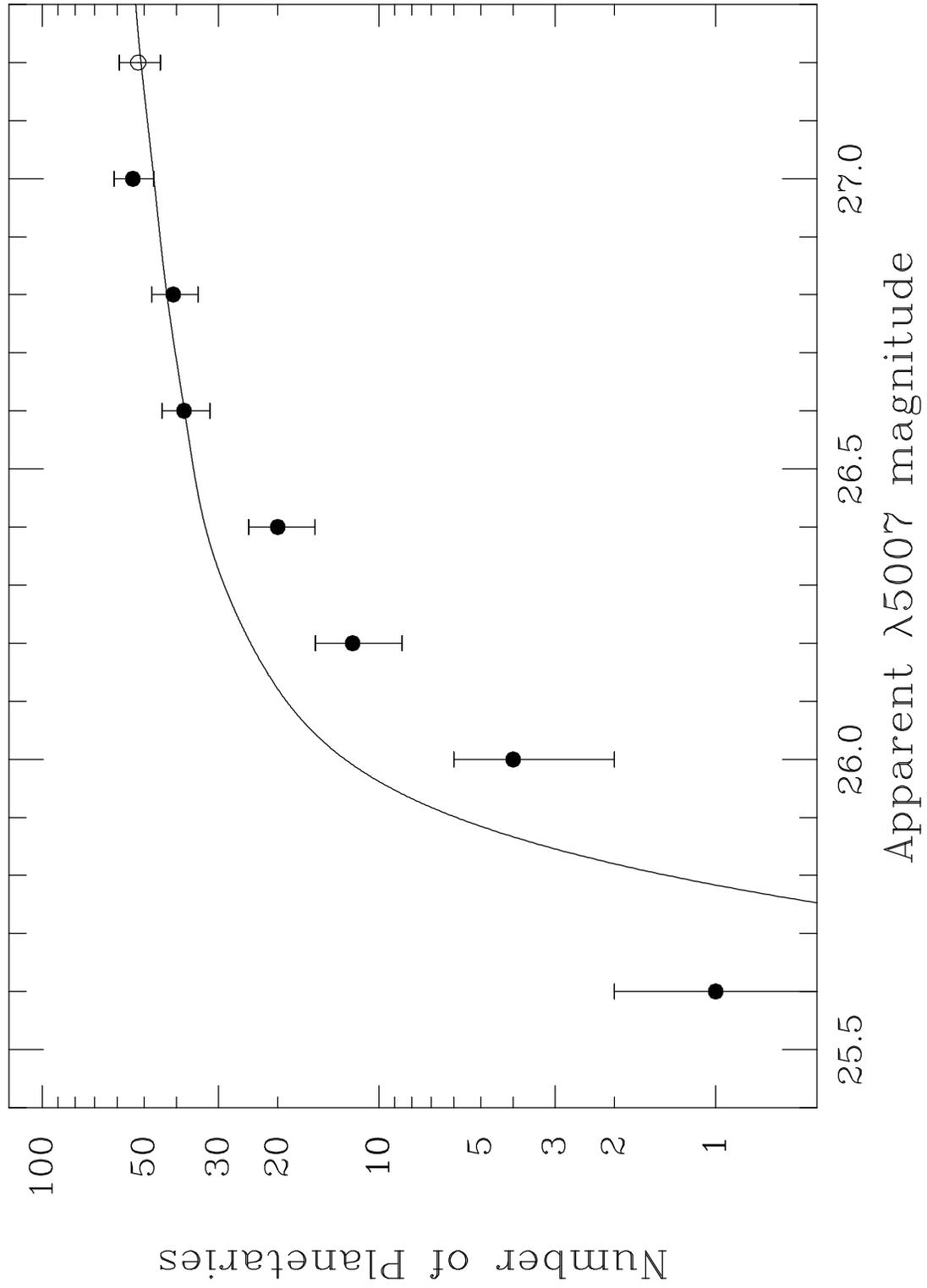
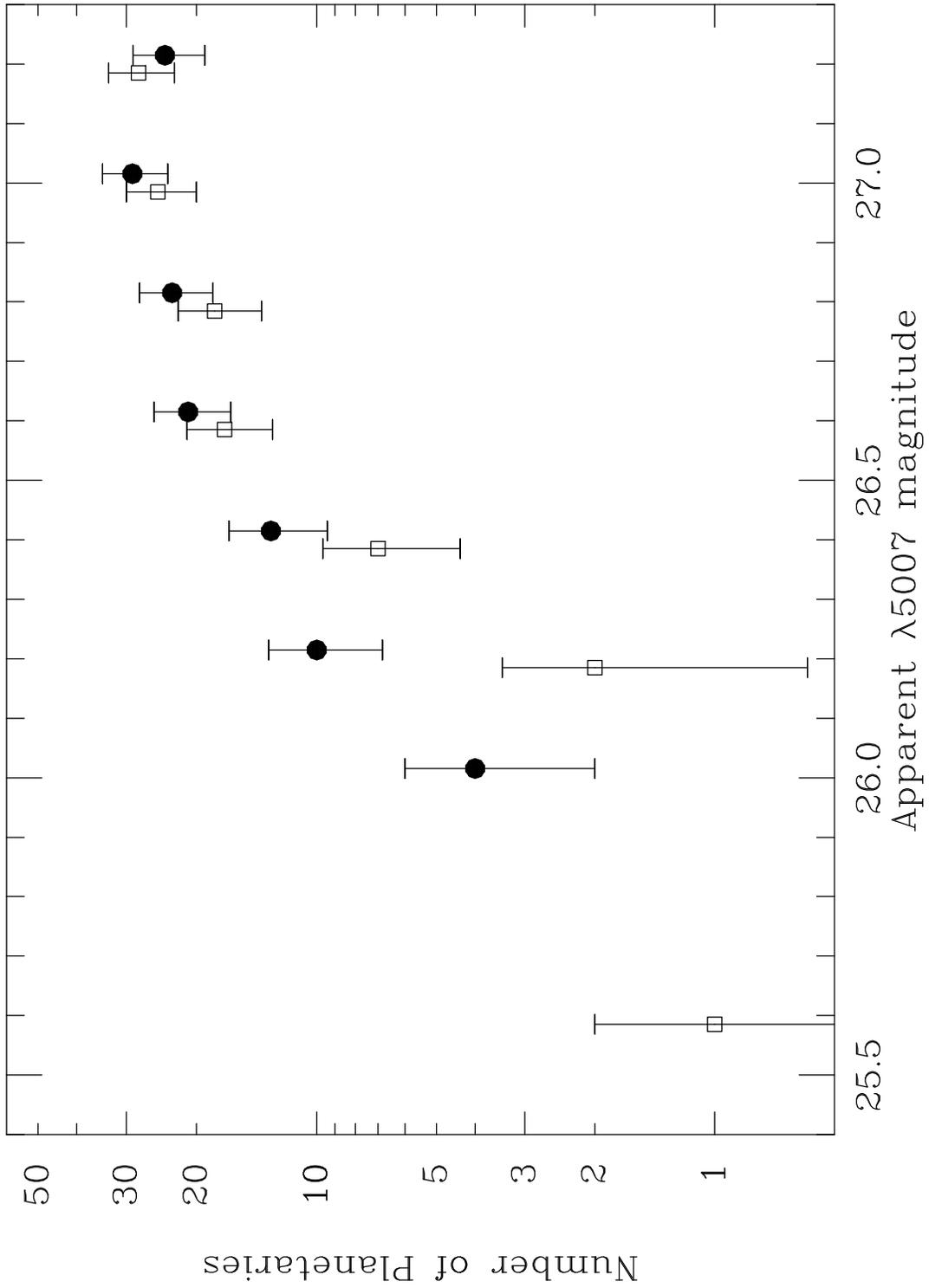
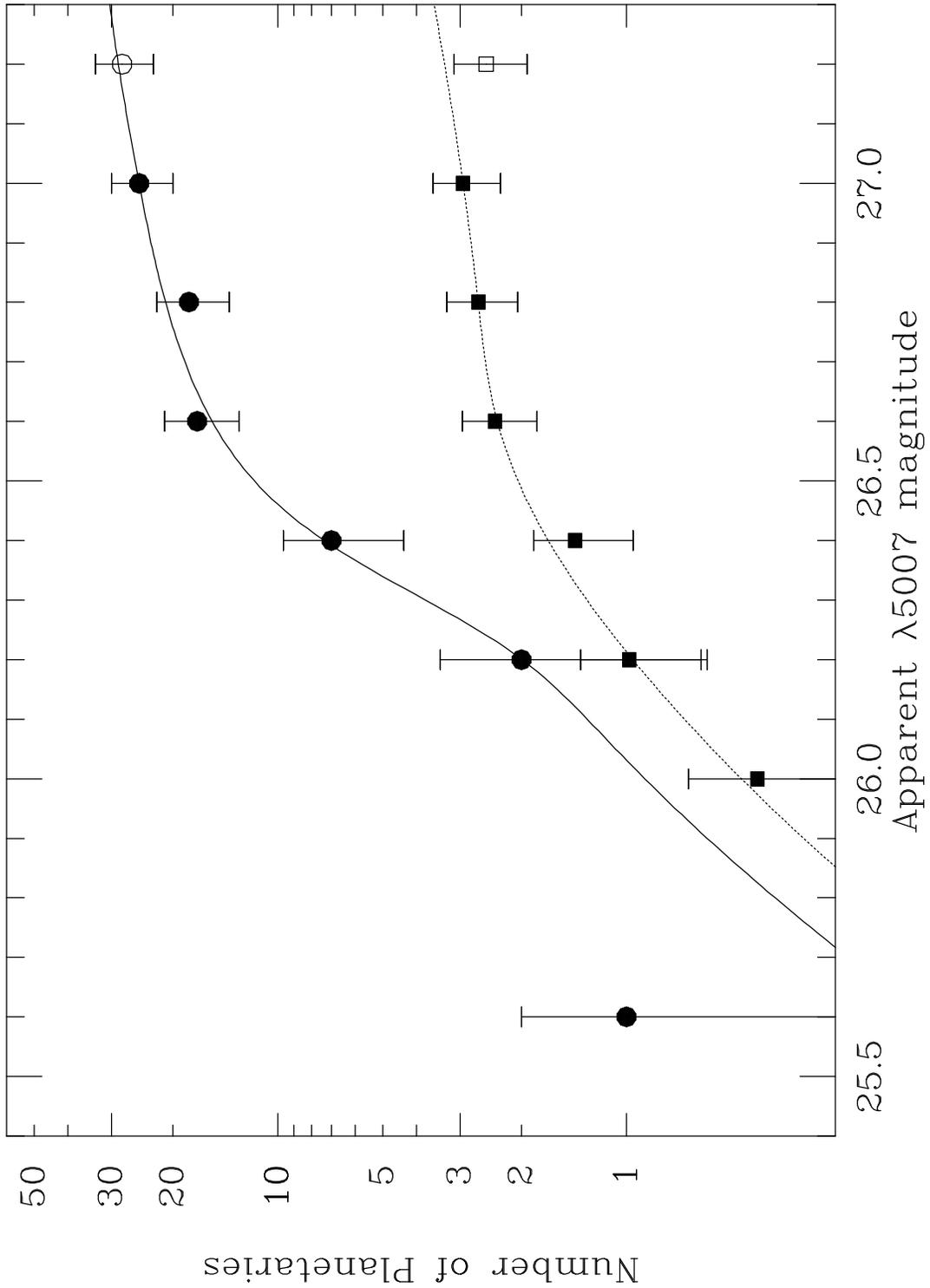
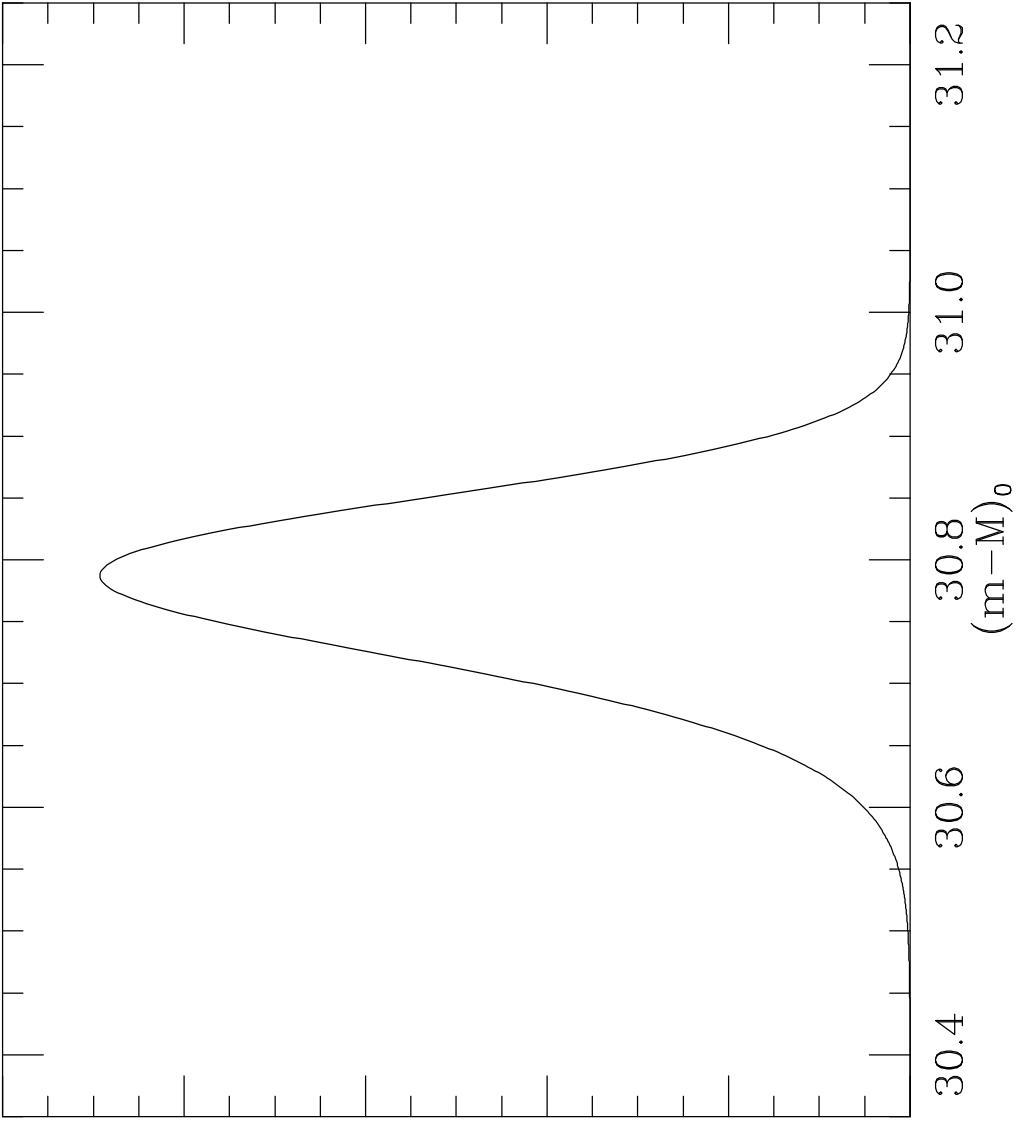
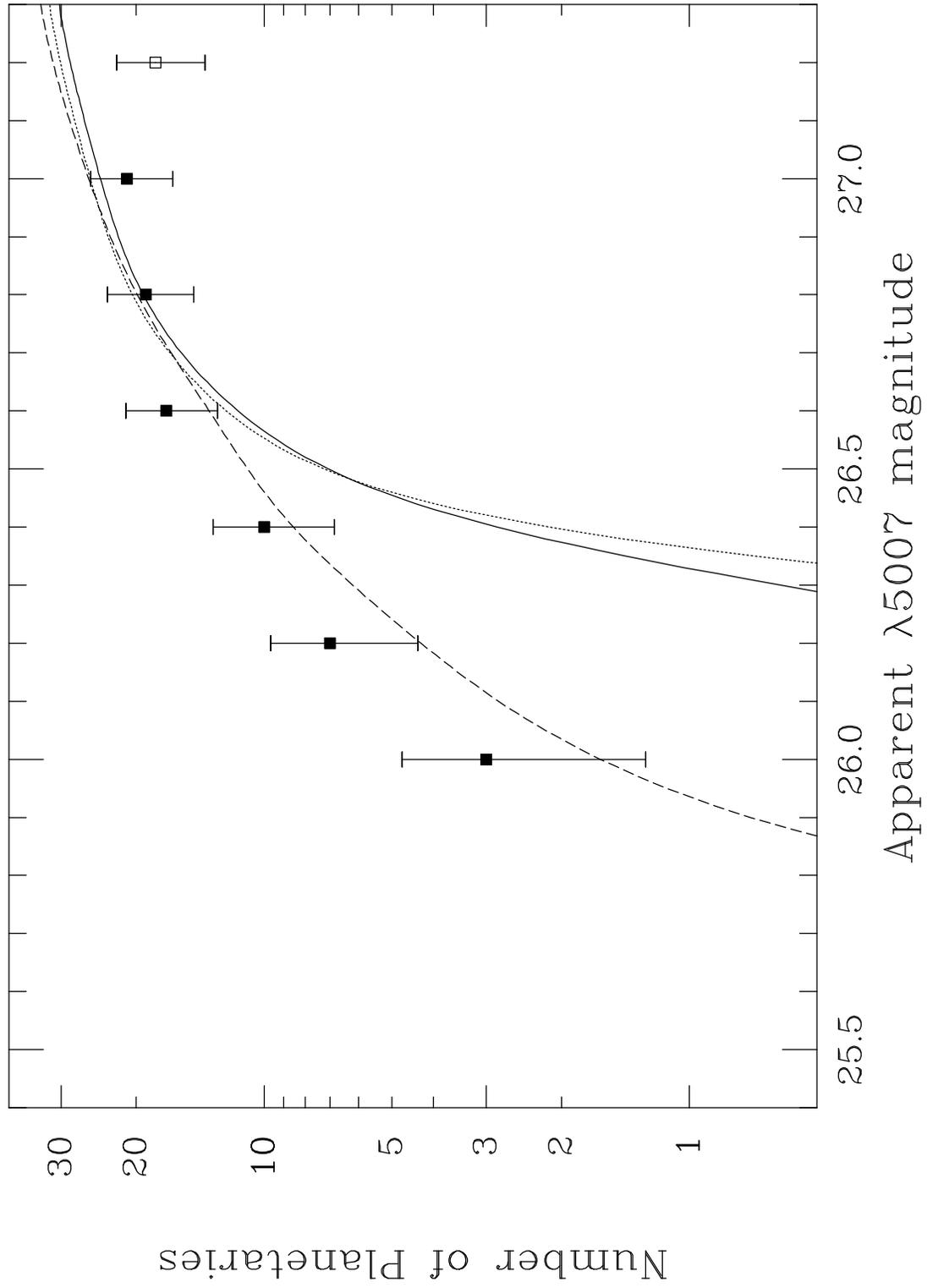
\end{document}